\documentclass[journal]{IEEEtran}
\IEEEoverridecommandlockouts
\usepackage{cite}
\usepackage{amsmath,amssymb,amsfonts}
\usepackage{algorithmic}
\usepackage{graphicx}
\usepackage{textcomp}
\usepackage{xcolor}
\def\BibTeX{{\rm B\kern-.05em{\sc i\kern-.025em b}\kern-.08em
    T\kern-.1667em\lower.7ex\hbox{E}\kern-.125emX}}

\usepackage{lettrine}
\usepackage{afterpage}
\ifCLASSOPTIONcompsoc
\usepackage[caption=false,font=normalsize,labelfont=sf,textfont=sf]{subfig}
\else
\usepackage[caption=false,font=footnotesize]{subfig}
\fi

\usepackage{tikz}

\usepackage{scalerel}
\usepackage{tikz}
\usetikzlibrary{svg.path}
\definecolor{orcidlogocol}{HTML}{A6CE39}
\tikzset{
  orcidlogo/.pic={
    \fill[orcidlogocol] svg{M256,128c0,70.7-57.3,128-128,128C57.3,256,0,198.7,0,128C0,57.3,57.3,0,128,0C198.7,0,256,57.3,256,128z};
    \fill[white] svg{M86.3,186.2H70.9V79.1h15.4v48.4V186.2z}
                 svg{M108.9,79.1h41.6c39.6,0,57,28.3,57,53.6c0,27.5-21.5,53.6-56.8,53.6h-41.8V79.1z M124.3,172.4h24.5c34.9,0,42.9-26.5,42.9-39.7c0-21.5-13.7-39.7-43.7-39.7h-23.7V172.4z}
                 svg{M88.7,56.8c0,5.5-4.5,10.1-10.1,10.1c-5.6,0-10.1-4.6-10.1-10.1c0-5.6,4.5-10.1,10.1-10.1C84.2,46.7,88.7,51.3,88.7,56.8z};
  }
}
\newcommand\orcidicon[1]{\href{https://orcid.org/#1}{\mbox{\scalerel*{
\begin{tikzpicture}[yscale=-1,transform shape]
\pic{orcidlogo};
\end{tikzpicture}
}{|}}}}
\usepackage{hyperref} %<--- Load after everything else

\newcommand{\HBF}{ {\text{HBF}}  }
\newcommand{\ABF}{ {\text{ABF}}  }

\newcommand{\thh}[1]{ {#1^{\text{th} } } }
\newcommand{\complexset}[2]{ \mathbb{C}^{#1 \times #2}  }
\newcommand{\complex}{  \mathbb{C}  }  
\newcommand{\diagonal}{ {\text{diag}}  }  
\newcommand{\expectation}{ \mathbb{E}  } 

\newcommand{\tx}{ {\text{T}}  }   
\newcommand{\rx}{ {\text{R}}  }   
\newcommand{\RF}{ {\text{RF}}  }  
\newcommand{\BB}{ {\text{BB}}  }  
\newcommand{\comm}{ {\text{c}}  }  
\newcommand{\rad}{ {\text{r}}  } 
\newcommand{\target}{ {\text{t}}  } 
\newcommand{\SI}{ {\text{SI}}  }
\newcommand{\IUI}{ {\text{IUI}}  }
\newcommand{\CRP}{ {\text{CRP}}  }
\newcommand{\W}{ \mathbf{W}  }
\newcommand{\w}{ \mathbf{w}  }

\newcommand{\cfreq}{ {\text{c}}  }
\newcommand{\Ltx}{ L_\tx  }
\newcommand{\Lrxrad}{ L_{\rx } }
\newcommand{\LRFtx}{ L^\RF_\tx  }    
\newcommand{\LRFrxrad}{ L^\RF_{\rx }  } 

\newcommand{\U}{ U  } 
\newcommand{\WRFtx}{ \W^\RF_\tx  } 
\newcommand{\WRFrxrad}{ \W^\RF_{\rx}  } 
\newcommand{\WBB}{ \W^\BB  } 
\newcommand{\wBB}{ \w^\BB  } 
\newcommand{\wRF}{ \w^\RF  } 
\newcommand{\wRFtx}{ \w^\RF_\tx  } 
\newcommand{\wRFrxrad}{ \w^\RF_{\rx}   } 
\newcommand{\xbold}{ \mathbf{x}  } 
\newcommand{\ybold}{ \mathbf{y}  } 
\newcommand{\vbold}{ \mathbf{v}  } 
\newcommand{\Hbold}{ \mathbf{H}  } 
\newcommand{\Bbold}{ \mathbf{B}  }
\newcommand{\Dbold}{ \mathbf{D}  }
\newcommand{\Nbold}{ \mathbf{N}  }
\newcommand{\Abold}{ \mathbf{A}  } 
\newcommand{\abold}{ \mathbf{a}  } 
\newcommand{\thetabold}{ \boldsymbol{\theta}  } 
\newcommand{\cbold}{ \mathbf{c}  }
\newcommand{\Cbold}{ \mathbf{C}  }
\newcommand{\eye}{ \mathbf{I}  }
\newcommand{\Fbold}{ \mathbf{F}  }
\newcommand{\Zerobold}{ \mathbf{0}  }

\newcommand{\Ktarget}{ K_\target  }
\newcommand{\SCS}{ \Delta f  }
\newcommand{\antSep}{ d_\text{ant}  }
\newcommand{\Nfreq}{ \mathcal{N}_\text{freq}  }
\newcommand{\Nang}{ \mathcal{N}_\text{ang}  }
\newcommand{\kHz}{ \text{kHz}  }
\newcommand{\MHz}{ \text{MHz}  }
\newcommand{\GHz}{ \text{GHz}  }
\newcommand{\dBi}{ \text{dBi}  }
\newcommand{\dB}{ \text{dB}  }
\newcommand{\dBm}{ \text{dBm}  }
\newcommand{\meter}{ \text{m}  }
\newcommand{\maxText}{ \text{max}  }

\begin{document}
\title{Beamformer Design and Optimization for Full-Duplex Joint Communication and Sensing\\at mm-Waves
%\title{Beamformer Design for Full-Duplex \\ Joint Communication and Sensing at mm-Waves
%\title{Beamforming Design for Hybrid and Analog Joint Communication and Sensing Systems at mm-Waves
%\thanks{This paper is an expanded version from the IEEE International Conference on Communications (ICC) 2020 (Wireless Communications Symposium), Dublin, Ireland, June 7-11, 2020 \red{[REF]}.}
\thanks{{This work was partially supported by the Academy of Finland (grants \#315858, \#319994, \#328214, \#338224, and \#346622), Nokia Bell Labs, and the Doctoral School of Tampere University. The work was also supported by Business Finland under the ``RF Convergence'' and the ``5G VIIMA'' projects.}}
\thanks{C. Baquero Barneto, T. Riihonen, S. D. Liyanaarachchi, M. Heino and M. Valkama are with the Unit of Electrical Engineering, Tampere University, Finland.}
\thanks{N. Gonz\'alez-Prelcic is with the Department of Electrical and Computer Engineering, North Carolina State University, Raleigh, NC 27606 USA.}
}

\author{\IEEEauthorblockN{
Carlos Baquero Barneto \orcidicon{0000-0002-0156-935X}\,, \IEEEmembership{Student Member, IEEE},
Taneli Riihonen \orcidicon{0000-0001-5416-5263}\,, \IEEEmembership{Member, IEEE},\\
Sahan Damith Liyanaarachchi \orcidicon{0000-0002-2852-854X}\,, \IEEEmembership{Student Member, IEEE},
Mikko Heino \orcidicon{0000-0003-3637-3477}\,, \IEEEmembership{Member, IEEE},\\
Nuria Gonz\'alez-Prelcic \orcidicon{0000-0002-0828-8454}\,, \IEEEmembership{Senior Member, IEEE}, and
Mikko Valkama \orcidicon{0000-0003-0361-0800}\,, \IEEEmembership{Senior Member, IEEE}}
}

\maketitle

\begin{abstract}
In this article, we study the joint communication and sensing (JCAS) paradigm in the context of millimeter-wave (mm-wave) mobile communication networks. We specifically address the JCAS challenges stemming from the full-duplex
%simultaneous transmit-receive (TX-RX) 
operation and from the co-existence of multiple simultaneous beams for communications and sensing purposes. To this end, we first formulate and solve beamforming optimization problems for hybrid beamforming based multiuser multiple-input and multiple-output JCAS systems. The cost function to be maximized is the beamformed power at the sensing direction while constraining the beamformed power at the communications directions, suppressing interuser interference and cancelling full-duplexing related self-interference (SI). We then also propose new transmitter and receiver beamforming solutions for purely analog beamforming based JCAS systems that maximize the beamforming gain at the sensing direction while controlling the beamformed power at the communications direction(s), cancelling the SI as well as eliminating the potential reflection from the communication direction and optimizing the combined radar pattern (CRP). Both closed-form and numerical optimization based formulations are provided. We analyze and evaluate the performance through extensive simulations, and show that substantial gains and benefits in terms of radar transmit gain, CRP, and SI suppression can be achieved with the proposed beamforming methods.
%compared to the state-of-the-art.
%platforms, enabling new opportunities and interesting applications.
%We investigate two different JCAS architectures, covering the two preferred design options at mm-wave frequencies: analog and hybrid analog-digital systems.
%Each architecture, presents different challenges, demanding alternative beamforming designs. The hybrid design enables multiple-input and multiple-output (MIMO) communications with multiple users while MIMO radar can be also implemented to sense the environment. In addition, we analyze the analog architecture, which demands more challenging beamforming techniques to efficiently suppress the interference between the communication and sensing systems. In addition, the presented JCAS beamformers for both architectures incorporate a novel spatial suppression scheme to cancel the wideband self-interference (SI) signal.
%Finally, numerical evaluations demonstrate that substantial gains and benefits can be achieved with the proposed beamforming methods.
\end{abstract}

% \begin{IEEEkeywords}
% 5G New Radio (NR), 6G, beamforming, full-duplex, joint communication and sensing, mm-wave, multibeam, radar, self-interference.
% \end{IEEEkeywords}


% ############################################################################
% INTRODUCTION
% ############################################################################

%\vspace{-4mm}
\section{Introduction}
\label{sec:Introduction}
% Ideas to discuss
% \begin{itemize}
%     \item Why we assume two different configurations? Explain properties and challenges of analog and hybrid digital-analog beamforming.
% \end{itemize}

\lettrine[lines=2]{\textbf{J}} \enskip OINT communication and sensing (JCAS) technology is experiencing a rapidly growing interest in both defense and civilian application domains, in response to the increasingly congested electromagnetic spectrum \cite{journal_1}. Another driving factor is the evolution of the wireless communication networks to higher frequency bands, where larger channel bandwidths and directional antenna systems allow for extracting accurate positioning and sensing information \cite{journal_1,journal_67}. In general, there are three main research areas reflecting different levels of convergence between the communication and radar/sensing functions or systems, namely the so-called \textit{co-existence}, \textit{cooperation} and \textit{co-design} paradigms\cite{journal_2}. 
%Firstly, \textit{coexistence} methods were adopted, where the communication and radar systems treat each other as interferers, sensing the spectrum to detect whether the desired frequency band is available. This approach is also commonly known as cognitive radio. 
%Secondly, \textit{cooperative} techniques where communication and sensing systems share certain information to effectively mitigate the mutual interference and thus achieve an improved spectral performance were introduced. 
In this work, we focus on the \textit{co-design} area, which seeks to merge the communication and sensing functionalities into the joint shared platform with common waveform and hardware \cite{journal_1, journal_2, journal_14, journal_23}. Such an approach can benefit both communications and sensing operations while relaxing the spectrum congestion, providing thus an efficient and appealing alternative compared to classical fully separated stand-alone systems \cite{journal_1}. Timely application fields contain, e.g., vehicular systems, unmanned aerial vehicles, residential security, building analytics and health monitoring \cite{myPaper_6_WCM20, myPaper_4_ICL20, journal_6,journal_36, journal_66}, to name a few.

In the context of cellular networks, integrating efficient sensing features into the fifth generation (5G) new radio (NR) and the future sixth generation (6G) mobile communication networks is one research area of largely increasing importance \cite{conference_49, journal_67}. Specifically, mobile networks are evolving to support operation at the so-called millimeter-wave (mm-wave) bands ($30$--$300~\GHz$), with the current 5G NR specifications reaching up to $52.6~\GHz$ \cite{3GPPTS38104}. %For that purpose, a new mm-wave frequency range FR2 ($24.25$-$52.6~\GHz$) will be combined with the traditional sub-$6~\GHz$ frequency range FR1 ($0.4$-$6~\GHz$).
It is expected that the beyond $52.6~\GHz$ bands will also be utilized, through later 5G NR standard releases, while in the 6G-era networks will be further expanding towards the sub-THz frequencies \cite{journal_67,journal_68,9318749}. 
%In \cite{journal_68}, the authors analyze the main physical layer challenges for future mobile communication systems at these frequency bands. 
%

From the sensing and JCAS perspectives, the transition to the mm-wave bands exhibits great opportunities and benefits. 
%The large bandwidths used for enhanced communications also improve the radar time/range resolution, i.e., the ability to distinguish very close targets \cite{myPaper_3_Asilomar19}. Due to the high path losses and directional antenna systems at the mm-waves, either line-of-sight (LoS) propagation or non-line-of-sight (NLoS) with a predominant reflection path is typically observed. This reduces the multipath components facilitating enhanced sensing operation and performance. In addition, high frequencies and subsequently small wavelengths imply more compact antennas which allow to implement large amounts of antenna units and thereon highly directive antenna arrays even in small form-factor devices, improving the angular observation capabilities in the systems \cite{journal_14, journal_54}.
The large bandwidths used at mm-waves enable receivers with enhanced ability to resolve the delays of the different multipath components, which results in better range estimation \cite{myPaper_3_Asilomar19}.   
In addition, high frequencies and subsequently small wavelengths imply more compact antennas, which allow to deploy large amounts of antenna units and thereon implement highly directive antenna arrays even in small form-factor devices, improving the angular observation capabilities of the systems \cite{journal_14, journal_54}. 
Finally, propagation is also different at mm-waves and above; due to limited diffraction and substantial scattering, mm-wave channels tend to have few dominant multipath components beyond a strong line-of-sight and a non-line-of-sight with a few reflections, enabling sparse channel models which are easier to relate to the propagation environment in order to obtain sensing information.

\subsection{State-of-the-Art}
In the recent literature, three main operation modes have been investigated to incorporate JCAS functionalities to wireless communication networks, such that the base station (BS) and/or the user equipment (UE) are utilized also for sensing purposes \cite{journal_3, journal_63}. 
The first JCAS mode is known as \textit{downlink active sensing}, and refers to the case where a BS collects the reflections stemming from its own downlink signal. In this case, the BS operates essentially as a monostatic radar where the sensing transmitter (TX) and receiver (RX) are colocated \cite{myPaper_2_TMTT19}. Alternatively, the BS can be used for \textit{downlink passive sensing} by collecting the reflections from downlink signals of different BSs, representing a passive radar scheme. Finally, the UE can be utilized for \textit{uplink sensing}, operating as a mobile JCAS system by collecting the reflections of its own transmit signal as a monostatic device \cite{journal_4}. Alternatively, the UE uplink signals can be exploited by the neighbouring BSs for passive sensing purposes. In general, these JCAS scenarios present different requirements and various design challenges as described in \cite{journal_63}. One of the most prominent research problems is the so-called joint waveform design, as the same waveform is shared between the communications and the sensing tasks. As most of the modern wireless networks are orthogonal frequency-division multiplexing (OFDM)-based, different OFDM variants and related optimization approaches have been recently studied in the JCAS context \cite{myPaper_2_TMTT19, sahan_TWC2021}.  
%In this work, we will focus on different design challenges for the \textit{downlink active sensing} case as it will be explained shortly, however, the same design principles presented in this paper can by incorporated in  \textit{downlink passive sensing} and \textit{uplink sensing} configurations.

% In this work we are going to focus on the main hardware design challenges

\begin{figure*}[t]
    \centering
        \subfloat[Analog beamforming JCAS architecture]{\includegraphics[width=0.5\textwidth]{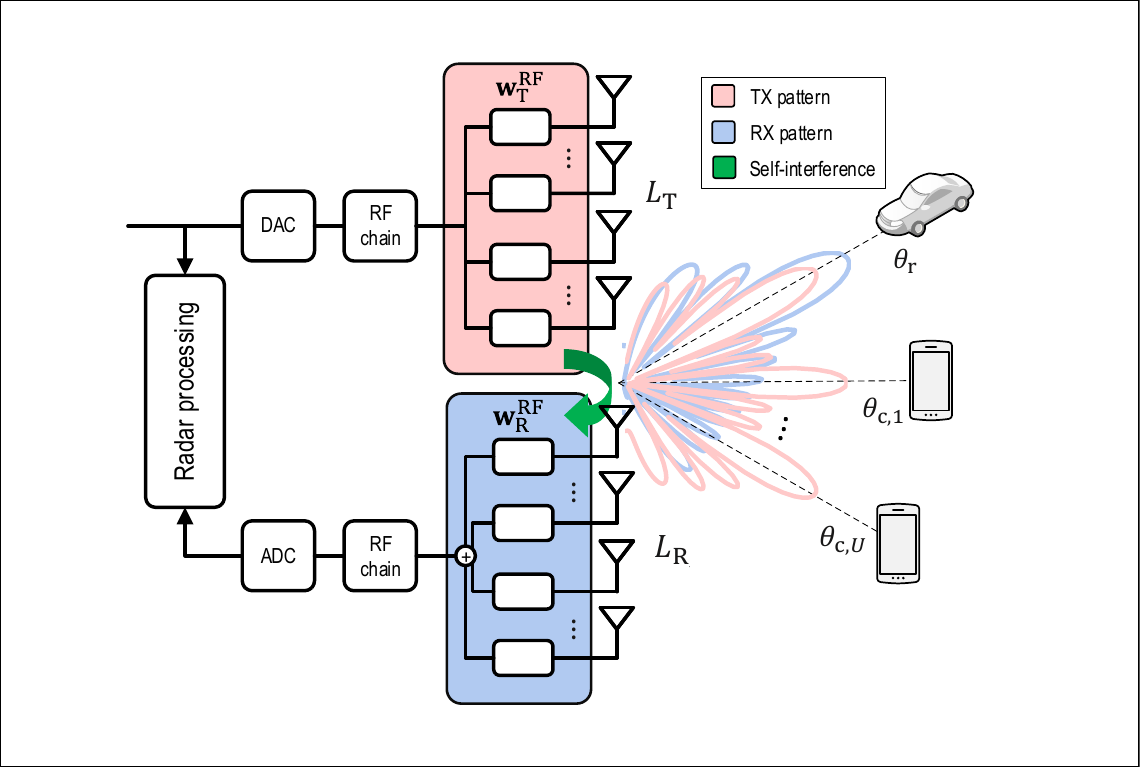}
        \label{fig:blockDiagram_analog}}
        \subfloat[Hybrid beamforming MU-MIMO JCAS architecture]{\includegraphics[width=0.5\textwidth]{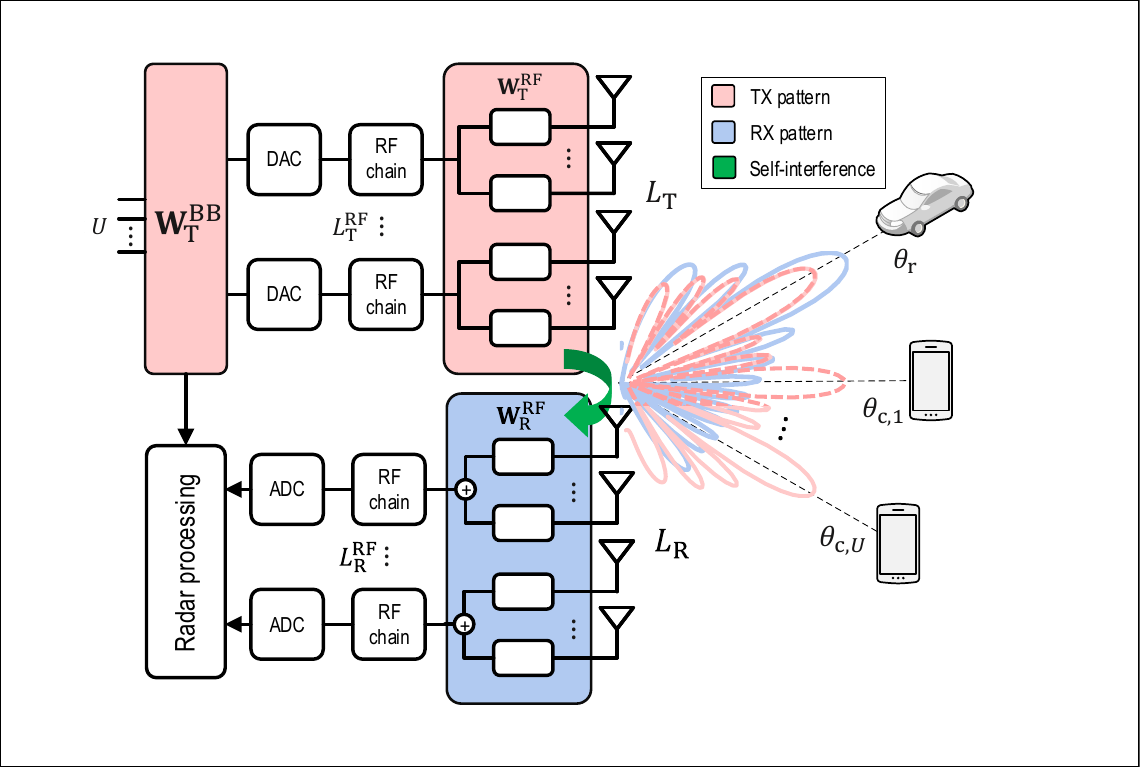}
        \label{fig:blockDiagram_MIMO}}
    \caption{Considered JCAS beamforming architectures with multiple communication beams for $U$ users at directions $\thetabold_{\comm} =[\theta_{\comm,1},\ldots,\theta_{\comm,U}]^T$ while simultaneously sensing the environment with a directional beam at $\theta_\rad$.
    Specifically, (a) illustrates the considered analog array based JCAS architecture while (b) shows the corresponding hybrid MU-MIMO JCAS architecture. }
    \label{fig:blockDiagram}
\end{figure*}

The recently studied mm-wave JCAS systems build commonly on massive antenna arrays with directive patterns to compensate for the small aperture of each antenna element
% for the associated large propagation losses 
and to obtain directional information of the targets \cite{journal_59, journal_55}. Depending on the implementation assumptions, these can be further categorized into analog, digital and hybrid analog--digital beamforming systems. In analog beamforming JCAS systems, a single transmit stream is subject to a set of analog weights and transmitted from a single RF chain for both communication and sensing purposes \cite{journal_7, journal_37}. 
In contrast, with digital beamforming and multiple TX/RX chains, the JCAS platform is able to perform multiple-input and multiple-output (MIMO) communication and sensing by transmitting multiple parallel streams through several RF chains. However, purely digital beamforming based approach, where each antenna element is connected to its own RF chain, is not necessarily implementation-feasible at mm-waves due to challenges related to power consumption, costs and silicon area \cite{journal_59} --- especially when the number of the antenna units is large. To overcome this problem, one can consider hybrid architectures where the overall MIMO beamforming task is split into the analog and digital domains.
In this work, we address the design and optimization challenges of both purely analog and hybrid analog-digital beamforming configurations in the JCAS context, as illustrated in Fig.~\ref{fig:blockDiagram}. 

While beamforming techniques have been extensively studied over the past decades for communications at large, some recent works have proposed new beamforming designs to satisfy the additional requirements stemming from the JCAS operation specifically. 
In \cite{journal_7, journal_37}, the so-called multibeam approach using steerable analog antenna arrays is identified as one of the main enablers for mm-wave JCAS systems. The proposed multibeam scheme provides separate controllable beams for communication and sensing with different design requirements.
In \cite{journal_35}, in turn, joint TX beamforming for multiuser (MU) MIMO communications and MIMO radar is investigated. In this case, the JCAS system transmits a sum of independent communication and sensing waveforms, forming multiple beams. 
The effects of beamforming weight quantization in multibeam JCAS systems are analyzed in \cite{journal_38}, focusing purely on the TX side and thus neglecting joint TX--RX optimization. 
%It is noted that these recent reference works focus only on the 
%This work only addressed the TX beamforming design, despite being the joint TX-RX beamforming optimization one key enabler of efficient JCAS operation.

%Enabling sensing functionalities in wireless communication systems demands challenging modifications. From the sensing perspective,
Finally, an important JCAS design and deployment challenge is related to the TX--RX isolation \cite{journal_2,myPaper_6_WCM20,9442375}. To this end, two duplexing modes can basically be considered known as half-duplex and full-duplex (FD) \cite{journal_9, journal_26}. In half-duplex sensing systems, such as ordinary pulse radars, the RX observes and collects the targets' reflections only when the TX has finalized the transmission of a sensing pulse. 
Therefore, the TX and RX never operate simultaneously, ensuring thus automatically high TX--RX isolation. However, such approach is not feasible for JCAS in wireless communication networks, where the transmit waveform is by default imposed by the communication requirements and the corresponding system specifications \cite{sahan_TWC2021, thesis_braun2014, conference_12, conference_13}. As a consequence, FD operation is required implying a simultaneous transmit-and-receive (STAR) mode \cite{journal_11} --- even in time-division duplexing (TDD) based networks, when viewed from the sensing function's point of view. 
Thus, one major technical challenge is the self-interference (SI) between TX and RX, which requires SI cancellation (SIC) to provide sufficient TX--RX isolation \cite{journal_31}. This is one of the JCAS challenges addressed in this article, through the beamforming design and optimization.

\subsection{Contributions, Novelty, and Organization}

In this article, we propose multiple new beamforming design and optimization solutions for mm-wave JCAS systems. We cover both analog array and hybrid MU-MIMO based JCAS scenarios, and propose novel TX and RX beamforming solutions to address particularly the challenges related to (\emph{i}) SI stemming from the STAR operation and (\emph{ii}) the co-existence and interference between the simultaneous communication and sensing beams. 
As a fundamental basis, we first provide comprehensive JCAS system models for OFDM-based networks, describing how TX and RX beamformer weights, beamforming architectures, multiple simultaneous beams, multiple targets and the SI contribute to the observed signal at sensing RX. In addition, gain pattern models are also derived, that together with the received signal models form the basis for the beamforming optimization. 
To this end, the main technical contributions and scientific novelty of this article can be summarized as follows:
\begin{itemize}
    %\item We provide comprehensive JCAS system models for OFDM-based networks, describing how TX and RX beamformer weights, beamforming architectures, multiple simultaneous beams, multiple targets and the SI contribute to the observed signal at sensing RX. 
    %
    %covering both JCAS platforms that simultaneously sense the environment with a directive and configurable beam while multiple beams are dedicated for communication functionalities. In this paper, we investigate two different JCAS architectures, covering the two preferred options at mm-wave frequencies: hybrid MU-MIMO JCAS architecture and analog array JCAS architecture, as illustrated in Fig.~\ref{fig:blockDiagram}.
    %This section presents the radar model including the SI challenge stemming from the required STAR operation. 
    % In addition, gain pattern models are also derived, that together with the received signal models form the basis for the beamforming optimization. 
    \item We formulate and solve TX and RX beamforming optimization problems for hybrid MU-MIMO JCAS systems that maximize the beamformed power at the sensing direction while constraining the beamformed power at the communication directions, suppress interuser interference (IUI) and cancel SI in a controlled manner.
    % We formulate and solve TX and RX beamforming optimization problems for hybrid MU-MIMO JCAS systems that allow to incorporate an additional beam for sensing while the communication requirements are met. In addition, these beamforming schemes suppress the interuser interference (IUI) and cancel SI in a controlled manner.
    \item We propose new TX and RX analog beamforming solutions for JCAS systems, in both closed-form and numerical optimization based formulations, which simultaneously provide multiple beams for communications and sensing, suppress the SI and cancel the undesired reflections stemming from the communication beams by optimizing the so-called combined radar pattern (CRP).
    %We propose new TX and RX analog beamforming solutions for JCAS systems, that maximize the beamforming gain at the sensing direction, controlling the beamformed power at the communications direction(s), suppressing the SI while also cancelling the potential reflection from the communication direction and optimizing the so-called combined radar pattern (CRP). Both closed-form and numerical optimization based formulations are provided.
    %
    %different TX and RX beamformer techniques for the considered analog and MIMO JCAS architectures. In Section~\ref{sec:MIMO_beamforming}, we propose a joint design of the analog and digital beamformers for the hybrid MU-MIMO JCAS architecture. In Section~\ref{sec:Analog_beamforming}, alternative beamforming methods are analyzed for the analog array JCAS architecture. In particular, we propose a novel technique to optimize the combined radar pattern (CRP) by designing the RX weights.
    %\item We present a novel beamformer approach for JCAS which simultaneously adopts a spatial suppression scheme to cancel the wideband SI signal and suppress the inter-user interference (IUI) in MU scenarios. The details are presented in Section~\ref{sec:NSP}.
    \item The performance of the proposed methods is assessed and mutually compared through comprehensive numerical evaluations, demonstrating that substantial gains and benefits can be achieved, in terms of radar TX  gain, CRP and SI suppression, compared to the existing reference approaches.
\end{itemize}

The rest of this article is organized as follows:
Section~\ref{sec:SystemModel} develops the fundamental system models.
%are derived and stated forming the basis for the actual beamforming design and optimization methods. 
Sections~\ref{sec:MIMO_beamforming} and ~\ref{sec:Analog_beamforming} present the proposed TX and RX beamforming solutions for the hybrid beamforming MU-MIMO and analog beamforming JCAS scenarios, respectively. 
Section~\ref{sec:numericalResults} provides a vast collection of numerical results and their analysis, while Section~\ref{sec:Conclusions} summarizes the conclusions.

%\blue{The rest of the paper is organized as follows...}

%JCAS systems requires simultaneous-transmit-and-receive

% In these works they only focus on the TX design. However, we think that in these JCAS system is crutial to consider both TX and RX optimization simultaneously to also address one of the main challenges of these systems, the SI

\textit{Notations:}
Vectors are denoted by bold lowercase letters (i.e., $\abold$), bold uppercase letters are used for matrices (i.e., $\Abold$) and scalars are denoted by normal font (i.e., $a$). The operators $(\cdot)^T$, $(\cdot)^*$, $(\cdot)^H$, $(\cdot)^\dagger$, $\expectation \{ \cdot \}$, $|\cdot|$ and $\|\cdot\|$ denote the transpose, conjugate, Hermitian transpose, pseudoinverse, expectation, absolute value and Euclidian norm, respectively.

% ############################################################################
% SYSTEM MODEL
% ############################################################################
\section{System Model}
\label{sec:SystemModel}

In this section, we formulate the fundamental signal and system models for the JCAS transceiver architectures depicted in Fig.~\ref{fig:blockDiagram}. Both analog beamforming (ABF) and hybrid beamforming (HBF) architectures are considered, as illustrated in the figure. Furthermore, from the communications point of view, a TDD-based network is assumed.
%which can simultaneously provide multiple communication links to downlink users while a directive beam senses the environment, as.

\subsection{Basic Assumptions and Notations}
% The JCAS system is assumed to provide $U$ communication beams at directions $\thetabold_{\comm} =[\theta_{\comm,1},\ldots,\theta_{\comm,U}]^T$ and one sensing/radar beam at direction $\theta_\rad$. 
The JCAS system is assumed to provide one sensing/radar beam at direction $\theta_\rad$ and $U$ communication beams at directions $\thetabold_{\comm} = [\theta_{\comm,1},\ldots,\theta_{\comm,U}]^T$. 
%In this paper, we assume two different JCAS architectures referred to as analog array JCAS architecture and hybrid MU-MIMO JCAS architecture as shown in Fig.~\ref{fig:blockDiagram}\subref{fig:blockDiagram_analog} and Fig.~\ref{fig:blockDiagram}\subref{fig:blockDiagram_MIMO}, respectively. These architectures present different beamformer design and implementation challenges, and thereby they will be treated separately in the rest of the paper. 
%
Furthermore, to facilitate some passive physical TX--RX isolation to begin with, separate TX and RX antenna systems are assumed, as illustrated in Fig.~\ref{fig:blockDiagram}. %, the considered JCAS system consists of two arrays used as TX and RX, respectively. 
For notational and conceptual simplicity, we assume uniform linear arrays (ULAs) for both TX and RX and focus on communications and sensing in the azimuth domain. However, the same design principles can be extended and applied also for the elevation domain and subsequently for  3D JCAS operation.
The TX system is used for both communication and sensing functionalities by sharing the same TX waveform, providing multiple beams, while the colocated RX system is executed simultaneously for sensing purposes, providing a single beam in the radar direction.
Due to the STAR operation, additional SI suppression is needed on top of the physical isolation, which in our work is pursued through SI channel knowledge and beamforming optimization.

The HBF JCAS architecture shown in Fig.~\ref{fig:blockDiagram}\subref{fig:blockDiagram_MIMO} %, we consider a hybrid analog-digital architecture where the beamforming process is divided into RF and baseband (BB) domains \cite{journal_59}. This configuration 
allows to transmit $U$ parallel streams, one for each communication user, through spatial multiplexing \cite{journal_59}. The TX side is assumed to contain a total of $\LRFtx$ RF chains and $\Ltx$ antenna elements, respectively, with $U< \LRFtx < \Ltx$; it is further assumed that each subarray is fed by one RF chain only. %In particular, we adopt a subarray configuration for both TX and RX sides, where each RF chain is connected to a subset of antennas. 
On the RX side, we assume a total of $\Lrxrad$ antenna elements, organized into subarrays and connected to $\LRFrxrad$ RF chains, with $\LRFrxrad < \Lrxrad$. 
%The outputs of these RF chains will be used for radar processing.
%
The ABF JCAS architecture shown in Fig.~\ref{fig:blockDiagram}\subref{fig:blockDiagram_analog} contains only a single RF chain on the TX and RX sides, i.e., $\LRFtx=\LRFrxrad=1$, serving all the antenna units. % are connected to the same RF chain. Moreover, it is important to highlight that with this architecture 
In this case, only a single spatial stream is feasible, however, multiple beams can still be capitalized through for example the frequency multiplexing of users in OFDM systems' context or alternatively, in broadcast type of transmission scenarios.
%transmitted for all the users, which implement a frequency-division multiplexing scheme to separate the desired OFDM subcarriers.

Similar to many reference works \cite{myPaper_2_TMTT19, sahan_TWC2021,thesis_braun2014, conference_12, conference_13}, OFDM is assumed as the baseline waveform. In the following, the OFDM subcarrier index is denoted by $n=0,\ldots,N-1$, with $N$ referring to the number of active subcarriers, and the signal models are expressed for an arbitrary subcarrier and multicarrier symbol duration in the baseband equivalent notion. % in the OFDM system. 
Furthermore, in HBF case, $x_{u,n}$, $u=1,\ldots,U$, denote the $U$ spatially multiplexed symbols at subcarrier $n$. In the ABF case, in turn, the corresponding data symbol is denoted by $x_{n}$. For simplicity, in the following system model we assume a static channel, however, the considered OFDM-based system model can be easily extended to time-varying channels and sensing of moving targets as shown in \cite{thesis_braun2014, conference_12, conference_13, myPaper_2_TMTT19}.

% Therefore, we assumme that each transmitted stream, $x_{u,n,m}$, consists of a grid of $N$ active subcarriers and $M$ OFDM symbols, where $n=0,\ldots,N-1$, $m=0,\ldots,M-1$ and $u=1,\ldots,U$ denote the subcarrier, symbol and communication user indices, respectively. Note that for the analog array JCAS case, we can omit the index $u$ from the transmitted stream $x_{n,m}$ as a single stream is shared by all the users. For simplicity, in the following system model we assume a static channel within the $M$ transmitted OFDM symbols and thus omit the index $m$ in the continuation. The considered OFDM-based system model can be easily extended to time-varying channels and sensing of moving targets as shown in \cite{thesis_braun2014, conference_12, conference_13, myPaper_2_TMTT19}. 

\subsection{Spatial TX Signals}
Considering first the HBF JCAS architecture, the beamformed spatial TX signal samples at subcarrier $n$ and within an arbitrary OFDM symbol index can be expressed as
\begin{align}
\label{eq:systemModel_radiatedSignalSum}
     \tilde{\xbold}^\HBF_{n} = \WRFtx \sum_{u=1}^{\U}\wBB_{\tx,u,n} x_{u,n},
\end{align}
where $\WRFtx \in \complexset{\Ltx}{\LRFtx}$ and $\wBB_{\tx,u,n}\in \complexset{\LRFtx}{1}$ are the TX RF weights and the frequency-dependent BB precoder weights for the $\thh{u}$ stream, respectively, while $x_{u,n}$, $u=1,\ldots,U$, denote the spatially multiplexed data symbols. Denoting
\begin{align}
    \WBB_{\tx,n} & = \left [\wBB_{\tx,1,n}, \wBB_{\tx,2,n}, \ldots, \wBB_{\tx,U,n} \right ], \\
    \xbold_{n} & = \left [x_{1,n}, x_{2,n},\ldots,x_{U,n} \right ]^{T},
    \label{eq:systemModel_TXstreams}
\end{align}
%where $\WBB_{\tx,n} \in \complexset{\LRFtx}{U}$ and $\xbold_{n} \in \complexset{U}{1}$ refer to the BB weights and the TX streams for all the users, respectively. The radiated
the spatial transmit signal in (\ref{eq:systemModel_radiatedSignalSum}) can be rewritten in vector--matrix notation directly as
\begin{equation} 
\label{eq:systemModel_RF_BB_TX}
    \tilde{\xbold}^\HBF_{n} = \WRFtx \WBB_{\tx, n} \xbold_{n}.
\end{equation}
In the ABF JCAS case, in turn, the corresponding TX signal reads
\begin{equation}
\label{eq:ABF_TX_signal}
    \tilde{\xbold}^\ABF_{n} = \wRFtx x_{n},
\end{equation}
where $\wRFtx \in \complexset{\Ltx}{1}$ stands for the TX RF weights applied for the single TX stream or TX sample $x_{n}$.

%In the rest of this section, we present the radar system model for both hybrid MU-MIMO and analog array JCAS architectures. Then, we put special emphasis on the SI channel and its estimation error. Finally, we derive the gain pattern for the JCAS system based on the TX and RX weights that will be the basis for the beamforming optimization.
For readers' convenience, the main system model variables are summarized in Table~\ref{tab:variableDefinition}. We next proceed with establishing the radar received signal models and the applicable gain pattern expressions that form the basis for the beamforming optimization in Sections III and IV.

\subsection{HBF JCAS: Radar Received Signal}

From the sensing perspective, the transmitted beamformed signal propagates over-the-air and interacts with multiple targets, producing reflections that will be observed by the radar RX. We assume $\Ktarget$ point targets with directions $\thetabold_\target= [\theta_{\target,1}, \ldots , \theta_{\target,\Ktarget}]^T$. 
The target reflections are modeled by the frequency-domain channel matrix ${\Hbold_{\target,n} = \diagonal(h_{\target,n,1}, \ldots, h_{\target,n,\Ktarget}) \in \complexset{\Ktarget}{\Ktarget}}$ where $\diagonal(\cdot)$ stands for the diagonal matrix with the argument entries $h_{\target,n,k}$ on the diagonal. For the $\thh{k}$ target reflection with $k=1,\ldots, \Ktarget$ and the $\thh{n}$ subcarrier, the target channel reads
\begin{align}
    h_{\target,n, k} & = b_{\target,n, k}     e^{-j2 \pi n \SCS \tau_{\target, k} } \text{ with} \nonumber \\
      b_{\target,n,k}   &= \sqrt{\frac{\lambda_{n}^2 \sigma_{\target,k}}
    {\left (  4\pi\right )^3 d_{\target,k}^4}},
\end{align}
where $\tau_{\target, k}$, $\sigma_{\target,k}$, $d_{\target,k}$ and $b_{\target,n, k}$ denote the two-way propagation delay, the radar cross-section, the distance and the attenuation factor of the $\thh{k}$ target based on the well-known \textit{radar range equation} \cite{thesis_braun2014}. 
The variables $\lambda_{n}$ and $\SCS$ refer to the wavelength of the $\thh{n}$ subcarrier and the OFDM waveform subcarrier spacing, respectively.

Considering the monostatic radar operation and that the targets are in the far field, we can assume the angle-of-departure (AoD) and the angle-of-arrival (AoA) for each target to be approximately the same, $\theta_{\target,k}$.
% , as noted already in (\ref{eq:systemModel_radarChannel}). 
The distance of each reflection can, in turn, be approximated as $d_{\target,k} \approx \frac{\tau_{\target,k} c_0}{2}$ where $c_0$ is the speed of light. Additionally, for an ideal TX ULA with $\Ltx$ antennas with a separation $\antSep$ between neighboring antenna elements, the TX array steering vector can be expressed as
\begin{equation}
    \abold_{\tx,n}(\theta_{\target,k}) =
    \left [ 
    1, e^{j\Phi_n(\theta_{\target,k})}
    ,\ldots,
    e^{j(\Ltx-1)\Phi_n(\theta_{\target,k})}
    \right ]^{T},
\end{equation}
where $\Phi_n(\theta_{\target,k}) = 2\pi \frac{\antSep }{\lambda_n}\sin{(\theta_{\target,k})}$ is the electrical AoD for the $\thh{k}$ reflection. The RX array steering vector and the electrical AoA are obtained similarly.
Then, the TX and RX steering matrices at the target directions $\thetabold_\target$ can be defined as 
$\Abold_{\tx,n}(\thetabold_\target) = [\abold_{\tx,n}(\theta_{\target,1}), \ldots,\abold_{\tx,n}(\theta_{\target,\Ktarget})] \in \complexset{\Ltx}{\Ktarget}$ 
and $\Abold_{\rx,n}(\thetabold_\target) = [\abold_{\rx,n}(\theta_{\target,1}), \ldots,\abold_{\rx,n}(\theta_{\target,\Ktarget})] \in \complexset{\Lrxrad}{\Ktarget}$, respectively.

Furthermore, the TX--RX coupling is represented by the SI channel matrix $\Hbold_{\SI,n} \in \complexset{\Lrxrad}{\Ltx}$, with the entries $\{ \Hbold_{\SI, n} \} _{l_\rx,l_\tx}$, $l_\rx=1,\ldots,\Lrxrad$ and $l_\tx=1,\ldots,\Ltx$, modeling the coupling channels between the $\thh{l_\tx}$ TX and the $\thh{l_\rx}$ RX antenna units.
Therefore, the received spatial frequency-domain signal at the RX antenna elements can be expressed as 
\begin{equation}
\label{eq:systemModel_radarChannel}
    \tilde{\ybold}_{n}^\HBF= 
    % \tilde{\ybold}_{\rad,n} = 
    \underbrace{
    \left(
    \Abold_{\rx,n} (\thetabold_\target)
    \Hbold_{\target,n}
    (\Abold_{\tx,n} (\thetabold_\target))^H 
    +\Hbold_{\SI,n}
    \right)
    }_{\Hbold_{\rad,n}}
    \tilde{\xbold}_{n}^\HBF
    + \tilde{\vbold}_{n},
    % + \tilde{\vbold}_{\rad,n},
\end{equation}
where 
%$\Abold_{\rx,n} (\thetabold_\target) \in \complexset{\Lrxrad}{\Ktarget}$ and $\Abold_{\tx,n} (\thetabold_\target) \in \complexset{\Ltx}{\Ktarget}$ are the TX and RX steering matrices at the target directions $\thetabold_\target$. The
the noise vector $\tilde{\vbold}_{n} \in \complexset{\Lrxrad}{1}$ is assumed to be additive, white and Gaussian. In above, the effective multiantenna radar channel is represented by the channel matrix $\Hbold_{\rad,n} \in \complexset{\Lrxrad}{\Ltx}$ which incorporates both the actual target reflections and the SI leakage between the TX and RX antennas.

\renewcommand{\arraystretch}{1.25}
\begin{table}[t]
  \begin{center}
  \setlength{\tabcolsep}{4.5pt}
    \caption{\textsc{ System Model Notations and Variables}}
    \label{tab:variableDefinition}
    \begin{tabular}{|c || c | c |}
      \hline
      \textbf{Variable} & \textbf{Definition} & \textbf{Size} \\
      \hline 
      %\multicolumn{3}{c}{\textbf{General variables} }\\
      \hline
      $\Ltx$ & Number of TX antenna elements & $-$\\
      \hline
      $\Lrxrad$ & Number of RX antenna elements & $-$\\
      \hline 
      $\LRFtx$ & Number of TX RF chains & $-$\\
      \hline
      $\LRFrxrad$ & Number of RX RF chains & $-$\\
      \hline
    %   $\Ls$ & Number of streams per user & $-$\\
    %   \hline
      $\U$ & Number of users & $-$\\
      \hline
      $\theta_\rad$ & Radar beam direction & $-$\\
      \hline
      $\theta_{\comm,u}$ & Communication beam direction for $\thh{u}$ user & $-$\\
      \hline
      $\thetabold_{\comm}$ & Communication beam direction vector & $U \times 1$\\
      \hline
      $\WRFtx$ & TX RF weight matrix & $\Ltx \times \LRFtx$\\
      \hline
      $\WRFrxrad$ & RX RF weight matrix & $\Lrxrad \times \LRFrxrad$\\
      \hline
      $\WBB_{\tx,n}$ & TX BB weight matrix for $\thh{n}$ subcarrier & $\LRFtx \times  \U$ \\
      \hline
      $\wRFtx$ & TX RF weight vector & $\Ltx \times 1$ \\
      \hline
      $\wRFrxrad$ & RX RF weight vector & $\Lrxrad \times 1$ \\
      \hline
    \end{tabular}
  \end{center}
\end{table}

Finally, the spatial signal at the RX antenna elements is processed with the RX RF beamformer. This yields a frequency-domain beamformed received signal model of the form
%(\ref{eq:systemModel_radarChannel}) is combined in different RF chains as
\begin{align} 
\label{eq:systemModel_RXRFchainSignal}
    % \ybold_{\rad,n} =
     \ybold_{n}^\HBF = 
     (\WRFrxrad)^H 
     \tilde{\ybold}_{n}^\HBF
    %  \tilde{\ybold}_{\rad,n},
    = (\WRFrxrad)^H
    \Hbold_{\rad,n}
    \WRFtx \WBB_{\tx, n} \xbold_{n}
    %+ \tilde{\vbold}_{n}
    +{\vbold}_{n}
    ,
\end{align}
where $\WRFrxrad \in \complexset{\Lrxrad}{\LRFrxrad}$ refers to the radar RX RF weights and %${\vbold}_{n} \in \complexset{\LRFrxrad}{1}$ with 
${\vbold}_{n} = (\WRFrxrad)^H \tilde{\vbold}_{n} \in \complexset{\LRFrxrad}{1}$ denotes the beamformed noise vector. The beamformed spatial observed signal in (\ref{eq:systemModel_RXRFchainSignal}) feeds then the actual radar processing algorithms.

\subsection{HBF JCAS: Gain Patterns}
We next address and formulate the effective gain patterns for the considered HBF JCAS system. Considering the high carrier frequency and the subsequent wavelength being in the order of millimeters, we assume far-field operation as the distance of the radar targets satisfies the condition $d_{\target,k} \gg \frac{(\Ltx \antSep)^2}{\lambda_{n}}$ and $d_{\target,k} \gg \frac{(\Lrxrad \antSep)^2}{\lambda_{n}}$. Based on (\ref{eq:systemModel_RF_BB_TX}) and (\ref{eq:systemModel_radarChannel}), the effective TX gain pattern for the $\thh{u}$ stream can then be described as
\begin{equation}
\label{eq:gain_MIMO_TX}
    G_{\tx,u,n}(\theta) =  | (\abold_{\tx,n}(\theta))^{H}  \WRFtx \wBB_{\tx,u,n} |^2.
\end{equation}
In the beamforming optimization, described in Section~\ref{sec:MIMO_beamforming}, the BB weights of a particular user $u$ can be designed to guarantee a minimum gain of $G_{\tx,u,n}(\theta_{\comm,u}) > \mu_u$ at the corresponding user direction $\theta_{\comm,u}$, with the value of $\mu_u$ being imposed by the communication performance requirements. At the same time, the BB weights can be used to minimize the interuser interference (IUI) by suppressing the effective TX response at the other users' directions $G_{\tx,u,n}(\theta_{\comm,u'}) \approx 0$ with $u' \neq u$.

Similarly, based on (\ref{eq:systemModel_radarChannel}) and (\ref{eq:systemModel_RXRFchainSignal}), we express the RX gain pattern for the $\thh{l}$ RX RF chain as  
\begin{equation}
\label{eq:gain_MIMO_RX}
     G^\RF_{\rx,l,n}(\theta) =  
     |(\wRF_{\rx,l}) ^{H}  
    %  |(\wRF_{\rx,l}) ^{H}
     \abold_{\rx,l,n} (\theta) |^2 ,
\end{equation}
where $\wRF_{\rx,l} \in \complexset{(\Lrxrad/\LRFrxrad)}{1}$ and $\abold_{\rx,l,n} (\theta) \in \complexset{(\Lrxrad/\LRFrxrad)}{1}$ with $l = 1,\ldots, \LRFrxrad$ denote the RF weights and the RX steering vector for the $\thh{l}$ RX RF chain, respectively. 
Based on the subarray architecture shown in Fig.~\ref{fig:blockDiagram} and assuming that the antenna elements are equally distributed among all the RF chains, the RX RF weight matrix reads
\begin{equation}
    \WRFrxrad = 
    \begin{bmatrix}
\wRF_{\rx,1} & \Zerobold & \cdots & \Zerobold \\
 \Zerobold & \wRF_{\rx,2} & \Zerobold & \vdots \\
\vdots & \Zerobold& \ddots  & \Zerobold\\
\Zerobold & \cdots & \Zerobold  & \wRF_{\rx,\LRFrxrad}
\end{bmatrix} 	,
\end{equation}
where $\Zerobold$ denotes the zero vector. Additionally, the TX RF weights $\WRFtx$ are distributed conceptually similarly to account for the considered subarray configuration in the TX side.

\subsection{ABF JCAS: Radar Received Signal and Gain Patterns}
We next present the corresponding radar received signal model and gain pattern expressions for the analog beamforming JCAS architecture. To this end, assuming the TX spatial signal model in (\ref{eq:ABF_TX_signal}), the beamformed frequency-domain received signal at the sensing receiver can be expressed as
\begin{align} 
     y_{n}^\ABF = (\wRFrxrad)^H \underbrace{(\Hbold_{\rad,n}\tilde{\xbold}^\ABF_{n}
     + 
    \tilde{\vbold}_{n})
    }_{ \tilde{\ybold}_{n}^\ABF}
    = (\wRFrxrad)^H \Hbold_{\rad,n}
    \wRFtx  x_{n} + 
    v_{n},
    % v_{\rad,n},
\end{align}
% \begin{align} 
%      y_{n}^\ABF = (\wRFrxrad)^H \Hbold_{\rad,n}\tilde{\xbold}^\ABF_{n}
%      + 
%     v_{n}
%     = (\wRFrxrad)^H \Hbold_{\rad,n}
%     \wRFtx  x_{n} + 
%     v_{n},
%     % v_{\rad,n},
% \end{align}
where $\wRFtx \in \complexset{\Ltx}{1}$ and $\wRFrxrad \in \complexset{\Lrxrad}{1}$ are the TX and RX RF beamforming weights, %for the analog array JCAS architecture, 
$v_{n} = (\wRFrxrad)^H \tilde{\vbold}_{n} \in \complex$ is the beamformed noise sample and $\Hbold_{\rad,n}$ is the effective radar channel matrix defined along (\ref{eq:systemModel_radarChannel}). Similar to (\ref{eq:gain_MIMO_TX}) and (\ref{eq:gain_MIMO_RX}), the TX and RX gain patterns for the analog array JCAS architecture can now be expressed as
\begin{align}
    G^\RF_{\tx,n} (\theta)& =  | (\abold_{\tx,n}(\theta))^{H}  \wRFtx |^2 , \nonumber \\
    G^\RF_{\rx,n} (\theta)& =  | (\wRFrxrad)^H \abold_{\rx,n} (\theta)|^2 .
\end{align}

Furthermore, %for the analog array JCAS architecture we evaluate the overall multibeam radar performance by introducing a new concept referred to as the 
we define and adopt the so-called \textit{combined radar pattern} (CRP), which refers to the total equivalent gain pattern for the radar system \cite{myPaper_5_ICC20}. 
In the ABF JCAS context, where TX and RX beam-patterns differ from each other, the CRP allows the evaluation of the overall radar performance.
The CRP can be expressed by the multiplication of both the TX and RX gain patterns as
\begin{equation}
    G^\RF_{\CRP,n} (\theta) = G^\RF_{\tx,n} (\theta) G^\RF_{\rx,n} (\theta).
\end{equation}
This CRP concept will be utilized along the ABF JCAS beamforming optimization, as discussed in greater details in Section~\ref{sec:CPSLoptimization}.

\section{\texorpdfstring{Beamformer Design and Optimization:\\Hybrid MU-MIMO JCAS System}{Beamformer Design and Optimization: Hybrid MU-MIMO JCAS System}}
\label{sec:MIMO_beamforming}

In this section, we address the beamformer design and optimization challenge for the HBF JCAS systems. We leverage the so-called null-space projection (NSP) approach \cite{book_banerjee2014linear} for mitigating the SI and IUI phenomena, and construct solvable optimization problems to maximize the beamformed power of the sensing beam while controlling that of the communications beams according to the given communications constraints.

\subsection{Null-Space Projection for SI and IUI Suppression}
\label{sec:NSP}
%In this subsection, we analyze the NSP approach used for SIC and IUI suppression with special focus on the hybrid MU-MIMO JCAS architecture, although the same concepts can be applied for the analog case. From the SIC perspective, we aim to suppress 
Building on the signal model in (\ref{eq:systemModel_RXRFchainSignal}), the effective SI channel between the TX streams and the sensing RX RF chains can be first expressed as
\begin{equation}
\label{eq:NSP_effectiveSIchannel}
    {\Cbold}_{\SI,n} = (\WRFrxrad)^H  \Hbold_{\SI,n} \WRFtx \WBB_{\tx, n} ,
\end{equation}
with ${\Cbold}_{\SI,n} \in \complexset{\LRFrxrad}{U}$. Specifically, the elements $\{ {\Cbold}_{\SI,n} \}_{l,u}$ with $l=1,\ldots,\LRFrxrad$ and $u=1,\ldots,U$ represent the effective SI channel between the $\thh{u}$ user stream and the $\thh{l}$ RX RF chain at subcarrier $n$. 
%In particular, the NSP method is implemented at each RX subarray separately by designing its RF weights. Therefore, 
To suppress the SI in the $\thh{l}$ receive chain, the corresponding RX beamformer weights $\wRF_{\rx,l}$ should satisfy the following condition:
\begin{equation}
\label{eq:NSP_effectiveSIchannel_vector}
    \cbold_{\SI,n,l}= 
    (\wRF_{\rx,l})^H  
    % (\wRF_{\rx,l})^H  
    \Hbold_{\SI,n} \WRFtx \WBB_{\tx, n}  = \mathbf{0}^T ,
\end{equation}
where $\cbold_{\SI,n,l} \in \complexset{1}{U}$ represents the $\thh{l}$ row of the matrix in (\ref{eq:NSP_effectiveSIchannel}). While the above condition is expressed for a particular subcarrier $n$, it can be extended 
%The considered mm-wave JCAS system is characterized by implementing high-performance communication and sensing functionalities using large frequency bandwidths. 
%We can extend this condition 
%to simultaneously suppress multiple 
to cover $\Nfreq$ subcarriers over the whole transmission bandwidth and subsequently to suppress wideband SI signal. This is expressed as
\begin{align}
\label{eq:NSP_multipleFrequencies}
    % (\wRF_{\rx,l})^H 
    (\wRF_{\rx,l})^H &
    \underbrace{\left [
     \hat{\Hbold}_{\SI,{n}_{1}} \WRFtx \WBB_{\tx, {n}_{1}}
    ,\ldots,
     \hat{\Hbold}_{\SI,{n}_{\Nfreq}} \WRFtx \WBB_{\tx, {n}_{\Nfreq}}
    \right ]}_{=\Bbold_{l}} \nonumber
    \\
    &= (\wRF_{\rx,l})^H \Bbold_{l} =\mathbf{0}^T,
\end{align}
where ${n}_{1},\ldots, {n}_{\Nfreq}$ denote the subcarrier indices for which the NSP method is considered. Furthermore, as the multiantenna SI channel is not known by default, the corresponding SI channel estimate, denoted by $\hat{\Hbold}_{\SI, n} \in \complexset{\Lrxrad}{\Ltx}$, is utilized in the above expression. The impacts of imperfect SI channel estimation accuracy are addressed along the numerical results.

Similar to \cite{journal_12}, the NSP matrix can be deduced from (\ref{eq:NSP_multipleFrequencies})
based on the Moore--Penrose pseudoinverse approach ($\Bbold_{l} (\Bbold_{l})^\dagger \Bbold_{l} = \Bbold_{l}$), expressed as
\begin{equation}
\label{eq:NSP_NSPmatrix}
    \Nbold_{\SI,l} = (\eye - \Bbold_l (\Bbold_l) ^\dagger).
\end{equation}
Based on (\ref{eq:NSP_NSPmatrix}), we can thus reformulate the wideband NSP condition in (\ref{eq:NSP_multipleFrequencies}) as
\begin{equation}
    % (\tilde{\w}^\RF_{\rx,l})^H
    (\tilde{\w}^\RF_{\rx,l})^H
    \Nbold_{\SI,l}  \Bbold_{l}
    = \mathbf{0}^T,
\end{equation}
where ${\w}^\RF_{\rx,l} = (\Nbold_{\SI,l})^H \tilde{\w}^\RF_{\rx,l}$. In the actual beamforming design addressed in the next subsection, the auxiliary vector $\tilde{\w}^\RF_{\rx,l} \in \complexset{(\Lrxrad/\LRFrxrad)}{1}$ will be optimized. Finally, we define and quantify the average achieved SI suppression among the overall transmission bandwidth with a total of $N$ subcarriers as
\begin{equation}
    \bar{c}_{\SI,l,u} = \frac{1}{N}\sum_{n=0}^{N-1} \left | c_{\SI,n,l,u} \right |^2,
\end{equation}
where $c_{\SI,n,l,u} = \{ {\Cbold}_{\SI,n} \}_{l,u}$.

A similar NSP approach can be employed to also suppress the IUI in the hybrid MU-MIMO JCAS system. To this end, the TX BB weights of the $\thh{u}$ user can be used to minimize the IUI by suppressing the effective TX response at the other users' directions, i.e., $G_{\tx,u,n}(\theta_{\comm,u'}) \approx 0$ with $u' \neq u$. Based on (\ref{eq:gain_MIMO_TX}), the corresponding NSP condition for the IUI can now be expressed as
\begin{equation}
\label{eq:NSP_IUI_condition}
    \underbrace{
    \left [(\abold_{\tx,n}(\theta_{\comm,1}))^{H} \WRFtx, \ldots,
    (\abold_{\tx,n}(\theta_{\comm,U'}))^{H} \WRFtx \right ]
    }_{\Dbold_{u,n}}
    \wBB_{\tx,u,n} = \mathbf{0}.
\end{equation}
Finally, the TX BB weights are defined with the help of the corresponding NSP matrices $\Nbold_{\IUI,u,n} = (\eye - (\Dbold_{u,n})^\dagger \Dbold_{u,n})$ as
\begin{equation}
\label{eq:NSP_IUI_BB}
    \WBB_{\tx,n} = [\Nbold_{\IUI,1,n} \tilde{\w}^\BB_{\tx,1,n}, \ldots, 
    \Nbold_{\IUI,U,n} \tilde{\w}^\BB_{\tx,U,n}],
\end{equation}
where $\tilde{\w}^\BB_{\tx,u,n} \in \complexset{\LRFtx}{1}$ while ${\w}^\BB_{\tx,u,n} = \Nbold_{\IUI,u,n} \tilde{\w}^\BB_{\tx,u,n}$ denotes the auxiliary weight vector used along the beamforming optimization.

\subsection{Beamforming Optimization}
\label{sec:MIMO_beamforming_optimization}

The proposed hybrid MU-MIMO JCAS beamforming design jointly optimizes the TX RF and BB weights to maximize the TX power at the radar direction $\theta_\rad$. This beamformed power can be expressed as  
\begin{align}
\label{eq:Power_radar_MIMO}
    P_{\tx,\rad,n} &= 
    \expectation \{  | 
     (\abold_{\tx,n}(\theta_\rad))^{H} \WRFtx \WBB_{\tx, n} \xbold_{n} |^2   \} \nonumber \\
     & =  \sum_{u=1}^{U} \expectation \{   | 
     (\abold_{\tx,n}(\theta_\rad))^{H}  \WRFtx \wBB_{\tx,n,u} x_{u,n} |^2 \}  \\
     & =  \sum_{u=1}^{U}   
     G_{\tx,u,n}(\theta_\rad)    P_{u,n}, \nonumber
\end{align}
where $ P_{u,n} = \expectation \{ |x_{u,n} |^2\}$ and $G_{\tx,u,n}(\theta_\rad)$ denote the $\thh{u}$ user stream's TX power and the effective effective TX gain at the radar direction for the $\thh{n}$ subcarrier defined in (\ref{eq:gain_MIMO_TX}), respectively. 
%
%%%%%%%%%In general, when considering the 5G NR numerology and specifications \cite{3GPPTS38104}, the maximum channel bandwidth at the current mm-wave bands is $400~\MHz$. Therefore, if assuming a network deployment, e.g., at the 28~GHz band, the fractional bandwidth is only in the order of $1~\%$. This in turn implies that the beamformer optimization is approximately independent of the subcarrier index, and is thus formally pursued below at the subcarrier located at the center of the channel -- denoted with $n_\cfreq$.

For notational simplicity, the beamformer optimization is next formulated for the subcarrier located at the center of the channel --- denoted with $n_\cfreq$. To this end, the TX BB and RF weights, $\WBB_{\tx,n_\cfreq}$ and $\WRFtx$, are obtained  by defining and solving the following optimization problem:
\begin{subequations}
\label{eq:MIMO_opt_problem}
\begin{align}
    & \max_{\tilde{\W}^\BB_{\tx,n_\cfreq}, \WRFtx}   
    \, \sum_{u=1}^{U}   
     G_{\tx,u,n_\cfreq}(\theta_\rad)  \\ 
          \text{s.t.}    &\left \| \WRFtx \wBB_{\tx,u,n_\cfreq}  \right \| = 1, \; \forall u, \label{eq:MIMO_constraint1} \\
      &  G_{\tx,u,n_\cfreq}(\theta_{\comm,u}) \geq \mu_{u}, \; \forall u, \label{eq:MIMO_constraint2}\\
      &\WBB_{\tx,n_\cfreq} = [\Nbold_{\IUI,1,n_\cfreq} \tilde{\w}^\BB_{\tx,1,n_\cfreq}, \ldots, 
    \Nbold_{\IUI,U,n_\cfreq} \tilde{\w}^\BB_{\tx,U,n_\cfreq}] . \label{eq:MIMO_constraint3}
\end{align}
\end{subequations}
The effective TX powers are normalized by (\ref{eq:MIMO_constraint1}). In addition, the minimum effective TX gain for the $\thh{u}$ user at the specific user direction $\theta_{\comm,u}$ is constrained in (\ref{eq:MIMO_constraint2}), where $\mu_u$ is the required minimum gain imposed by the communication system and the corresponding performance requirements. 
Finally, in (\ref{eq:MIMO_constraint3}), the null-space projection structure in (\ref{eq:NSP_IUI_BB}) is imposed such that optimization is performed over the auxiliary BB weight matrix $\tilde{\W}^\BB_{\tx,n_\cfreq} = [\tilde{\w}^\BB_{\tx,1,n_\cfreq}, \ldots, \tilde{\w}^\BB_{\tx,U,n_\cfreq}] \in \complexset{\LRFtx}{U}$ to effectively suppress the IUI.

At the RX side, the goal is to maximize the RX beamforming gain at the radar direction $G^\RF_{\rx,l,n_\cfreq}(\theta_\rad)$ for each RF chain, while the SI is effectively suppressed through the NSP approach described in Section~\ref{sec:NSP}. Thus, the RF weights of the $\thh{l}$ RX RF chain are obtained by solving the following optimization problem:
\begin{subequations}
\label{eq:MIMO_RX_optimization}
\begin{align}
    & \max_{\tilde{\w}^\RF_{\rx,l}}   
    \,   | (\tilde{\w}^\RF_{\rx,l})^H \Nbold_{\SI,l} \abold_{\rx,l,n_\cfreq}(\theta_\rad) |^2 \\ 
     \text{s.t.}    &\left \|  \w^\RF_{\rx,l}  \right \| = 1, 
\end{align}
\end{subequations}
where ${\w}^\RF_{\rx,l} = (\Nbold_{\SI,l})^H \tilde{\w}^\RF_{\rx,l}$. This essentially leads to a spatial matched filter-like approach that is optimized through a proper phase alignment.
Furthermore, from the NSP matrix definition in (\ref{eq:NSP_NSPmatrix}), we can observe that $\Nbold_{\SI,l}$ is a Hermitian and idempotent matrix ($\Nbold_{\SI,l}=(\Nbold_{\SI,l})^H = \Nbold_{\SI,l} \Nbold_{\SI,l}$).
Thus, we can derive the optimal RX weights as
\begin{align}
\label{eq:MIMO_RX_weights}
    \w^\RF_{\rx,l} & = 
    \frac
    {(\Nbold_{\SI,l})^H \tilde{\w}^\RF_{\rx,l} }
    { \left \|  (\Nbold_{\SI,l})^H \tilde{\w}^\RF_{\rx,l}  \right \|}
    = \frac
    {(\Nbold_{\SI,l})^H \Nbold_{\SI,l} \abold_{\rx,l,n_\cfreq}(\theta_\rad) }
    { \left \| (\Nbold_{\SI,l})^H \Nbold_{\SI,l} \abold_{\rx,l,n_\cfreq}(\theta_\rad)  \right \|} \nonumber\\
    &=\frac
    {\Nbold_{\SI,l} \abold_{\rx,l,n_\cfreq}(\theta_\rad) }
    { \left \|  \Nbold_{\SI,l} \abold_{\rx,l,n_\cfreq}(\theta_\rad)  \right \|}.
\end{align}

\vspace{2mm}
\section{\texorpdfstring{Beamformer Design and Optimization: \\ Analog Array JCAS System}{Beamformer Design and Optimization: Analog Array JCAS System}}
\label{sec:Analog_beamforming}
We next propose and formulate beamforming design and optimization methods for the analog array JCAS architecture such that the selected beamformed communications requirements are met, while maximizing the ability to simultaneously sense targets in another direction. Both closed-form and numerical optimization based methods will be provided, while the main differences compared to the previous hybrid MU-MIMO system model are that only RF beamformers and one spatial stream per subcarrier are considered.

% For this purpose, the considered JCAS beamforming design needs to satisfy the following system requirements:
% \begin{itemize}
%     \item The TX RF weights need to provide multiple beams for $U$ communication users with directions $\thetabold_{\comm} =[\theta_{\comm,1},\ldots,\theta_{\comm,U}]^T$ while an additional radar beam at $\theta_\rad$ senses the environment.
%     \item In the RX side, the RX RF weights need to maximize the RX sensing gain at the corresponding radar direction $\theta_\rad$.
%     \item The RX weights are design to also cancel the possible clutter reflections stemming from the TX communication beams. 
%     \item The RX RF weights are also optimized to effectively cancel the strong SI stemming from the STAR operation by applying a NSP beamforming technique described in Section~\ref{sec:NSP}.
% \end{itemize}

\subsection{Closed-Form Solution}
\label{sec:analog_CF}

Building on the ABF JCAS signal and system models in Section~\ref{sec:SystemModel}, we provide a closed-form (CF) beamforming design procedure to define the TX and RX beamforming weights, $\wRFtx$ and $\wRFrxrad$, respectively. 
%In this method, the TX weights are designed independently to provide multiple beams for communication and sensing. Then, the RX weights are optimized accordingly to provide a single beam for sensing while also suppressing the TX--RX coupling channel.
At the TX side, we first generate separate weights for the corresponding communication and sensing beams. This can be achieved by choosing the TX RF weights to maximize the TX beamforming gain $G^\RF_{\tx,n_\cfreq} (\theta)$ at the given desired direction $\theta$, expressed formally as
\begin{subequations}
\begin{align}
    & \max_{\wRFtx(\theta)}   
    \,| (\abold_{\tx,n_\cfreq}(\theta))^{H}  \wRFtx (\theta)|^2 \\ 
     \text{s.t.}    &\left \|  \wRFtx (\theta) \right \| = 1.  
\end{align}
\end{subequations}
This yields the following normalized weights
\begin{equation}
\label{eq:CF_TX_separate}
    \wRFtx (\theta)= \frac{\abold_{\tx,n_\cfreq}(\theta) }
    {\left \| \abold_{\tx,n_\cfreq}(\theta) \right \|}.
\end{equation}

Then, similar to \cite{journal_7, myPaper_7_Asilomar20, journal_38}, the communication and sensing weights are combined coherently. Considering the $U$ communication beams and a single sensing beam, all coexisting simultaneously, the combined TX weights can be expressed as
\begin{equation}
\label{eq:CF_TX_combination}
    \wRFtx = \sqrt{\rho_\rad} \wRFtx (\theta_\rad)
    + \sum^{U}_{u=1} \sqrt{\rho_{\comm,u}} \wRFtx (\theta_{\comm,u}),
\end{equation}
where $\wRFtx (\theta_\rad)$ and $\wRFtx (\theta_{\comm,u})$ are the individually optimized TX RF weights for sensing and communications, respectively. The $\rho$-parameters control the energy distribution between the beamforming weights and subsequently the overall communication and radar performance. In particular, $\rho_\rad$ and $\rho_{\comm,u}$, with $\rho_\rad + \sum^{U}_{u=1} \rho_{\comm,u}=1$, control the energy of the sensing and the $\thh{u}$ communication beams, respectively. The obtained TX weights $\wRFtx$ are further normalized as $\left \| \wRFtx \right \| = 1$.

We then formulate a generalized optimization problem for designing the RX beamforming weights similar to (\ref{eq:MIMO_RX_optimization}), expressed as
\begin{subequations}
\begin{align}
    & \max_{\tilde{\w}^\RF_{\rx}}   
    \,   | (\tilde{\w}^\RF_{\rx})^H \Nbold_{\RF} \abold_{\rx,n_\cfreq}(\theta_\rad) |^2 \\ 
     \text{s.t.}    &\left \|  \wRFrxrad \right \| = 1, 
\end{align}
\end{subequations}
where $\wRFrxrad = (\Nbold_{\RF})^H \tilde{\w}^\RF_{\rx}$ and $\Nbold_{\RF} = (\eye - \Fbold \Fbold ^\dagger)$. %It is noted that the RX RF weights of the analog array JCAS architecture correspond to the specific case of (\ref{eq:MIMO_RX_optimization}) when $\LRFrxrad=1$.
Furthermore, compared to (\ref{eq:MIMO_RX_optimization}), we extend the considered NSP approach to simultaneously cancel the impact of the communication beam direction at the RX pattern. Based on (\ref{eq:NSP_multipleFrequencies}) and (\ref{eq:NSP_IUI_condition}), we thus generalize the NSP matrix incorporating $\Nfreq$ frequency nulls to suppress the SI and $\Nang$ angular nulls to cancel the reflections due to the communication beams, expressed formally as
\begin{equation}
\label{eq:analog_CF_freqAngNulls}
    \Fbold = [
    \underbrace{
    \hat{\Hbold}_{\SI,n_{1}} \wRFtx , \ldots, \hat{\Hbold}_{\SI,n_{\Nfreq}} \wRFtx 
    }_{\Nfreq \text{ frequency nulls}}
    ,\underbrace{
    \abold_{\rx,n_\cfreq}(\theta_1), \ldots ,
    \abold_{\rx,n_\cfreq}(\theta_{\Nang})
    }_{\Nang \text{ angular nulls}}
    ].
\end{equation}
%
%Similar to (\ref{eq:MIMO_RX_weights})
Finally, the RX RF weights for the analog array JCAS architecture are normalized as
\begin{equation}
    \w^\RF_{\rx} = \frac
    {\Nbold_{\RF} \abold_{\rx,n_\cfreq}(\theta_\rad) }
    { \left \|  \Nbold_{\RF} \abold_{\rx,n_\cfreq}(\theta_\rad)  \right \|}.
\end{equation}

For presentation clarity, we summarize the different RX beamforming configurations in Table~\ref{tab:CF_configurations}. The basic reference scheme from \cite{journal_7}, denoted by CF-A, does not pursue SI nor communication beam suppression. In this case, the RX weights are just optimized to provide a radar beam by $\Nbold_{\RF} = \eye$. The second configuration, denoted by CF-B, incorporates $\Nfreq$ frequency nulls in the NSP matrix definition to effectively suppress the SI signal. Finally, the third configuration CF-C is the most general and sophisticated, including $\Nfreq$ frequency nulls and $\Nang$ angular nulls to suppress the SI and the communication beam interference, respectively.

\begin{table}[t]
  \begin{center}
  \setlength{\tabcolsep}{4.5pt}
    \caption{\textsc{Different configurations for the RX beamformer in ABF JCAS systems, where $\Nfreq$ and $\Nang$ denote the numbers of frequency bins and angular directions for which the SI and communication beam impacts are nulled, respectively.  }}
    \label{tab:CF_configurations}
    \begin{tabular}{|c || c | c | c |}
      \hline
      \textbf{} & \textbf{CF-A} &  \textbf{CF-B} & \textbf{CF-C} \\
      \hline 
      \hline
      SI suppression & -- & $\mathcal{N}_\text{freq}$ & $\mathcal{N}_\text{freq}$\\
      Communication beam suppression & -- & -- & $\mathcal{N}_\text{ang}$\\
      \hline
    \end{tabular}
  \end{center}
\end{table}

\subsection{CPSL-based Optimization Solution}
\label{sec:CPSLoptimization}

The beamforming solution described in Section~\ref{sec:analog_CF} addresses both the SI leakage as well as the communication beam interference in the ABF JCAS context. However, to further improve the capability to suppress the communication beam interference in terms of, e.g., clutter from the environment, we next pursue a numerical optimization based beamformer design approach. Specifically, we formulate an optimization problem to design the RX RF beamforming weights that minimize the negative effects of the TX communication beams, by optimizing the combined peak sidelobe level (CPSL) of the CRP, while also efficiently suppressing the harmful SI leakage. To this end, a constrained least-squares problem is formulated as
\begin{subequations}
\label{eq:analog_PSLoptimization}
\begin{align}
    & \min_{\tilde{\w}^\RF_{\rx}}   
    \,   \frac{1}{S} \sum^{S}_{s=1} \eta_s
    \left | G^\RF_{\CRP,n_\cfreq} (\theta_s) - \tilde{G}^\RF_{\CRP,n_\cfreq} (\theta_s)\right|^2 \\ 
     \text{s.t.}    &\left \|  \wRFrxrad  \right \| = 1,  %\nonumber
\end{align}
\end{subequations}
where $\wRFrxrad = (\Nbold_{\RF})^H \tilde{\w}^\RF_{\rx}$ and the NSP matrix $\Nbold_{\RF}$ implements $\Nfreq$ frequency nulls as described in (\ref{eq:analog_CF_freqAngNulls}). The main goal of the optimization is to provide a CRP $G^\RF_{\CRP,n_\cfreq} (\theta_s)$ that is as close to the desired ideal CRP $\tilde{G}^\RF_{\CRP,n_\cfreq} (\theta_s)$, in least-squares sense, at the defined directions $\theta_s$ with $s =1,\ldots,S$. %, by minimizing the sum of the squares of the error residuals as shown in (\ref{eq:analog_PSLoptimization}). 
Additional weights $\eta_s$, $s =1,\ldots,S$, are allowed in the optimization to control and emphasize the error at specific directions.
We note that the above problem is convex and can be efficiently solved by numerical tools \cite{journal_27}.

\begin{figure}[t]
        \centering
        \includegraphics[width=1\columnwidth]{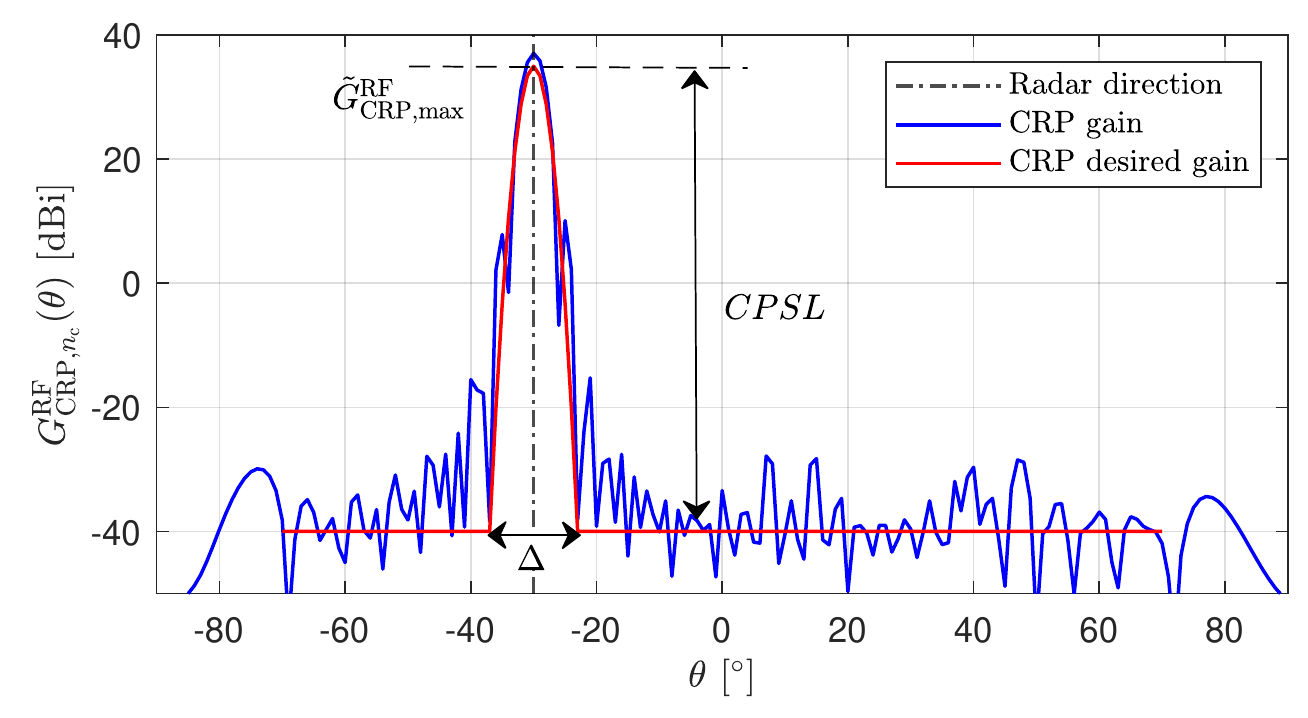}
        \vspace{-7mm}
        \caption{Illustration of a CRP with directional radar beam at $\theta_\rad=-30^{\circ}$. The proposed RX RF beamforming design seeks to provide a CRP that approximates well the desired gain pattern with $\tilde{G}^\RF_{\CRP,\maxText}$, $\Delta$ and $CPSL$ denoting the maximum gain of the CRP at the radar direction, the mask's width around the radar direction and the desired CPSL of the CRP, respectively.
        }
        \label{fig:PSL_optimization}
\end{figure}

In general, the CRP needs to provide a single beam at the sensing direction while the CPSL is minimized at the rest of directions. In this work, we approximate the desired CRP gain as the following piecewise parabolic function 
\begin{align}
    \tilde{G}^\RF_{\CRP,n_\cfreq} (\theta_s) = 
    \left\{ \begin{array}{cc} 
                \frac{ (\theta_s - \theta_\rad)^2 CPSL }{\Delta^2 /4} + \tilde{G}^\RF_{\CRP,\maxText}  & %\hspace{5mm} 
                \text{if $\left |  \theta_s - \theta_\rad \right | \leq  \frac{\Delta}{2}$}\\
                \tilde{G}^\RF_{\CRP,\maxText}  + CPSL & \text{if $\left |  \theta_s - \theta_\rad \right | >  \frac{\Delta}{2}$} \\
    \end{array} \right.
\end{align}
where $\tilde{G}^\RF_{\CRP,\maxText}$, $\Delta$ and $CPSL$ denote the maximum gain of the CRP at the radar direction, the mask's width around the radar direction and the desired CPSL of the CRP, respectively.
Figure~\ref{fig:PSL_optimization} presents an illustrative example of the CPSL optimization approach showing the main design parameters.

\begin{figure}[t]
    \centering
        \subfloat[Effective TX beam-patterns of the two streams]{\includegraphics[width=1\columnwidth]{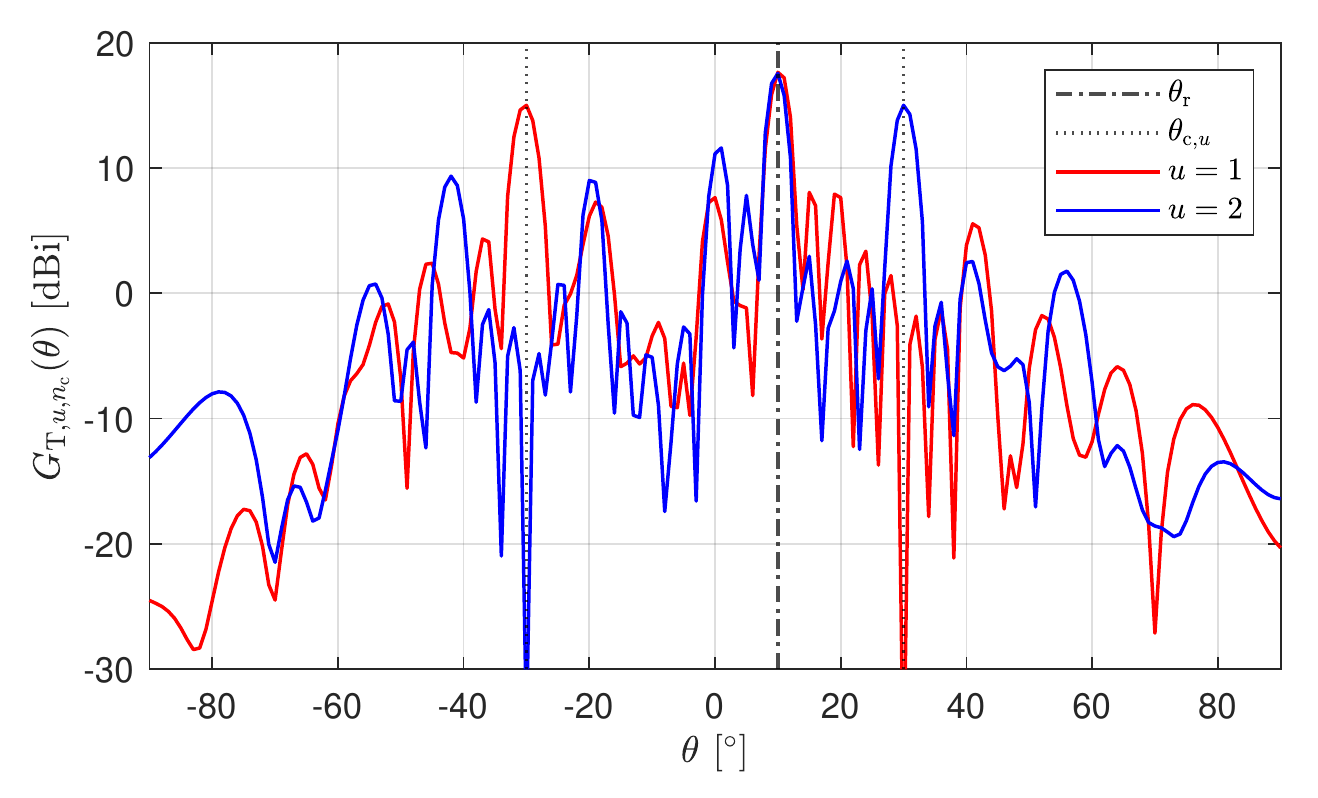}
        \label{fig:MIMO_TX}}
        \hfil
        \subfloat[RX beam-patterns of the four subarrays]{\includegraphics[width=1\columnwidth]{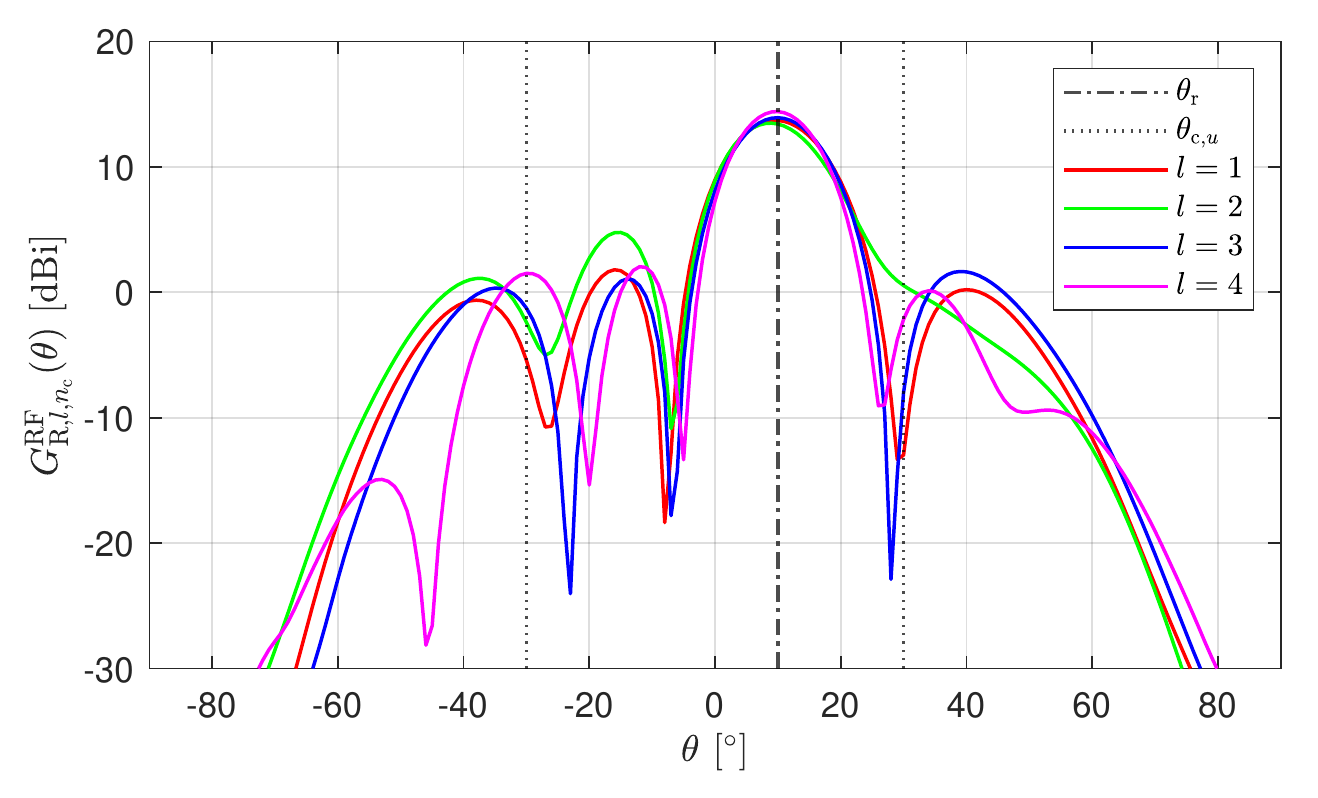}
        \label{fig:MIMO_RX}}
    \caption{Illustration of the relevant beam-patterns in a hybrid MU-MIMO JCAS system with a radar beam at ${\theta_\rad=10^\circ}$ and $U=2$ communication users located at $\theta_{\comm,1}=-30^\circ$ and $\theta_{\comm,2}=30^\circ$, respectively.
    The rest of the system parameters are ${\Ltx=\Lrxrad=32}$, $\LRFtx=8$, $\LRFrxrad=4$, $\mu_1 = \mu_2 = 15~\dBi$, and $\Nfreq=2$.}
    \label{fig:MIMO_patterns}
\end{figure}

\begin{figure}[t]
        \centering
        \includegraphics[width=1\columnwidth]{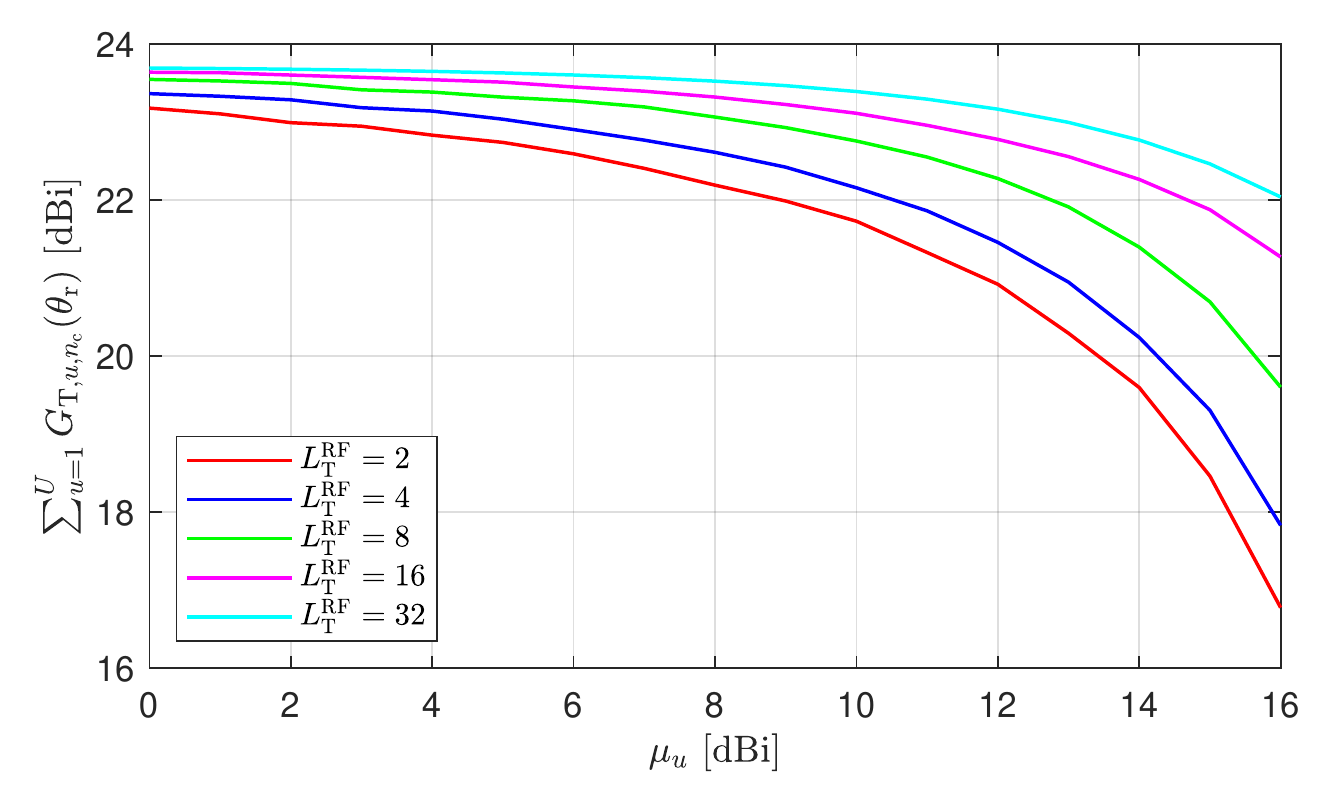}
        \caption{Illustration of the achievable beamformed radar TX gain in hybrid MU-MIMO JCAS context for varying communication TX gain $\mu_u$ and different numbers of TX RF chains $\LRFtx$. In this analysis, we assume the following parameters of $U=2$, ${\theta_\rad=10^\circ}$, $\theta_{\comm,1}=-30^\circ$, $\theta_{\comm,2}=30^\circ$, $\mu_1 =\mu_2 = \mu_u$ and ${\Ltx=32}$. The specific case of $\LRFtx=\Ltx=32$ corresponds to full digital TX beamforming. 
        }
        \label{fig:MIMO_muAnalysis}
\end{figure}

\begin{figure}[!t]
    \centering
        \subfloat[TX beam-pattern \cite{journal_7}]{\includegraphics[width=1\columnwidth]{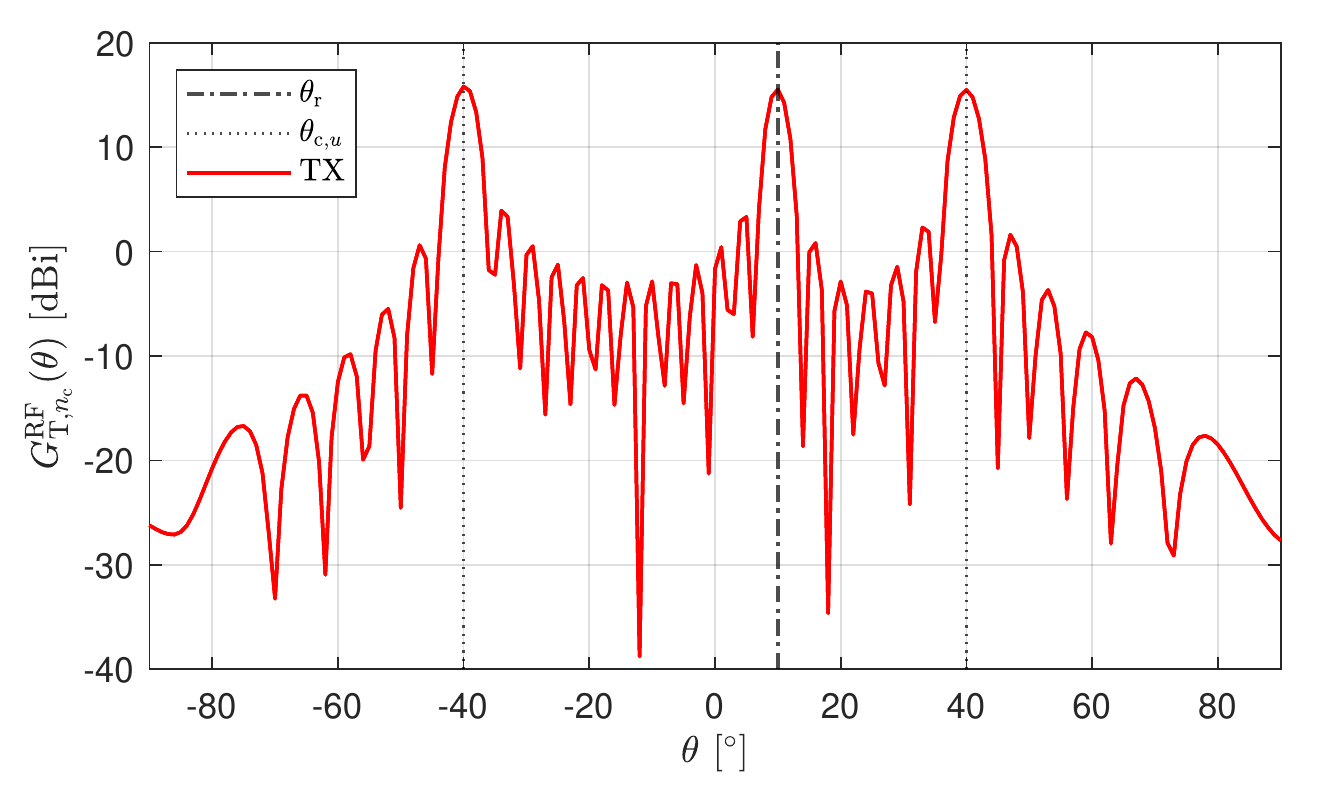}
        \label{fig:analog_TXpattern}}
        \hfil
        \subfloat[RX beam-patterns]{\includegraphics[width=1\columnwidth]{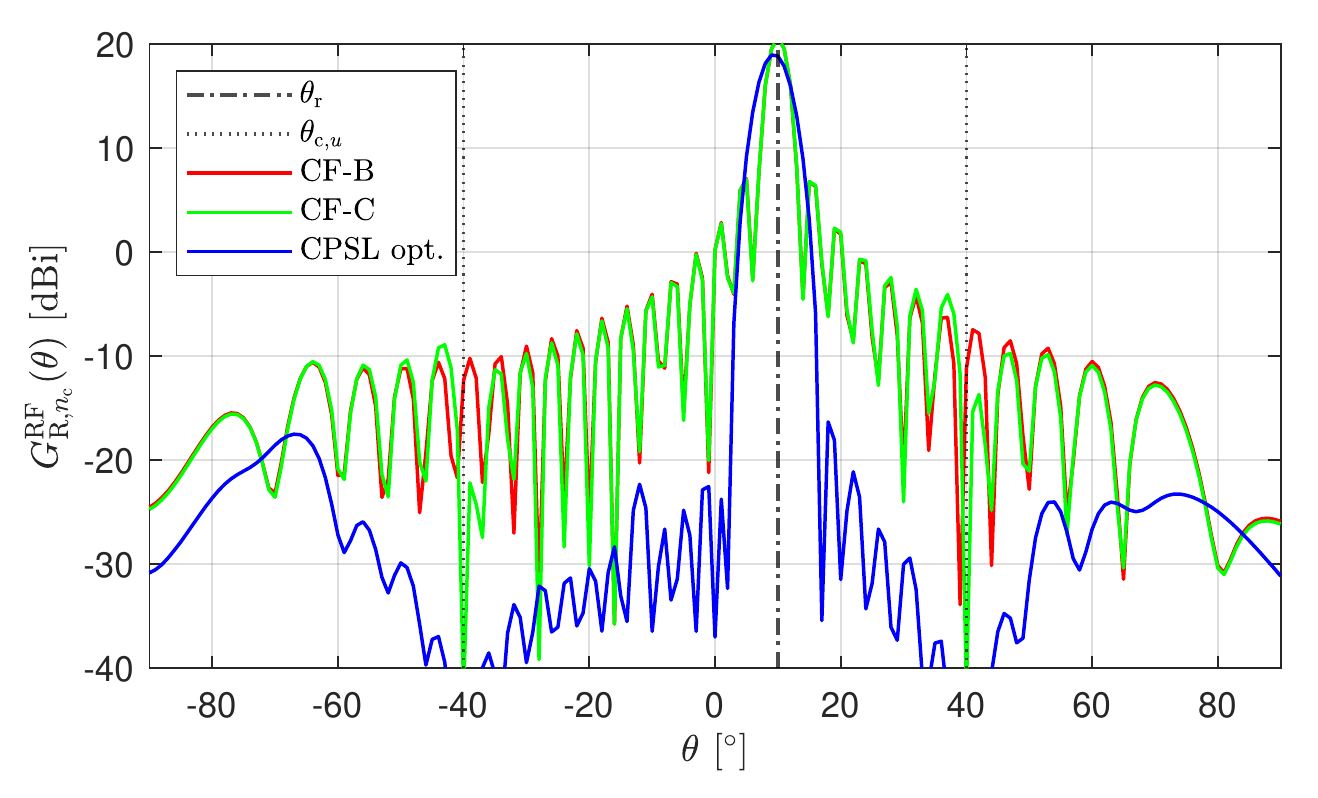}
        \label{fig:analog_RXpattern}}
        \hfil
        \subfloat[Combined radar beam-patterns]{\includegraphics[width=1\columnwidth]{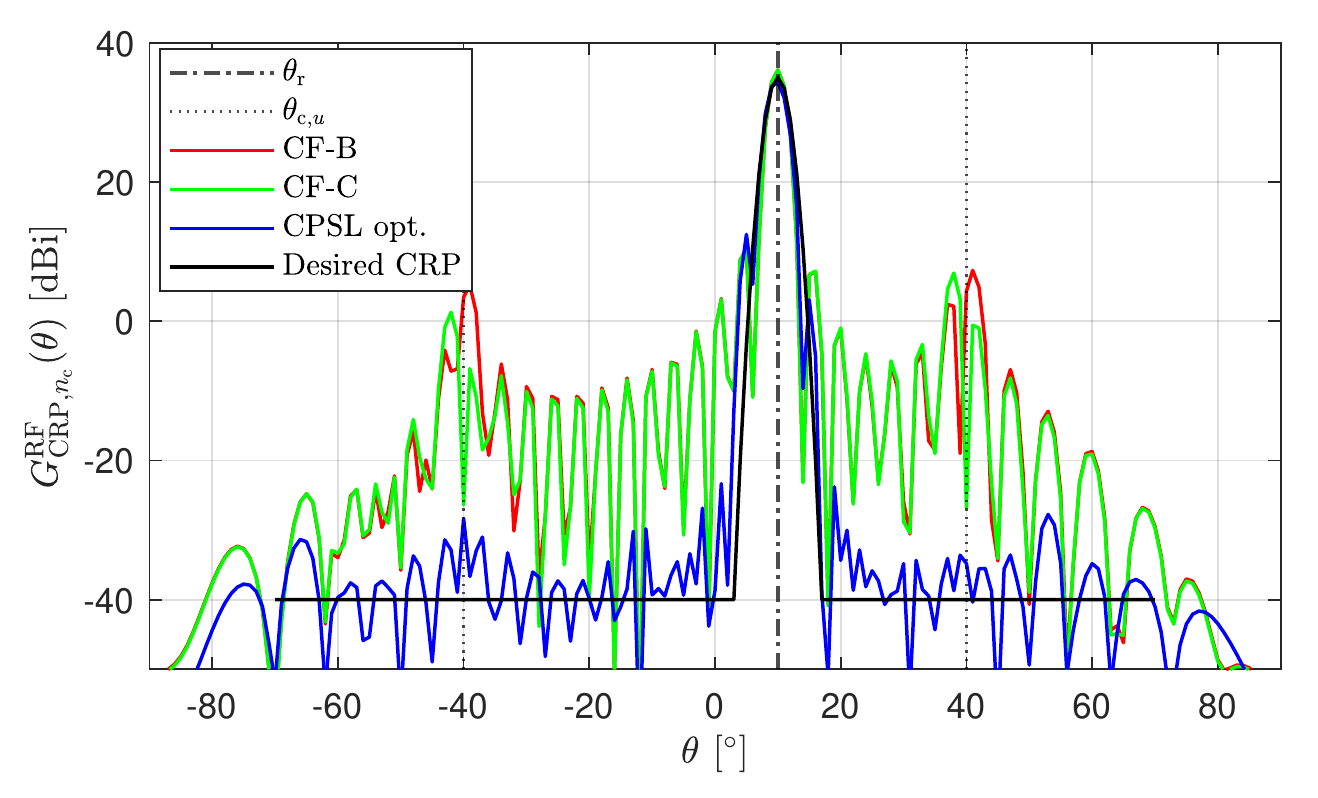}
        \label{fig:analog_CRPpattern}}
    \caption{Illustration of the beam-patterns with analog array JCAS architecture, with a radar beam at ${\theta_\rad=10^\circ}$ and $U=2$ communication users located at $\theta_{\comm,1}=-40^\circ$ and $\theta_{\comm,2}=40^\circ$, respectively. At the TX side, three beams are designed with the same relative energy share of $\rho_\rad=\rho_{\comm,1}=\rho_{\comm,2}=1/3$. In the CPSL optimization case, a desired response with $\tilde{G}^\RF_{\CRP,\maxText}=35~\dBi$, $\Delta=14^\circ$ and $CPSL=-75~\dB$ is used as shown in (c).
    The rest of the evaluation parameters are ${\Ltx=\Lrxrad=32}$, $\LRFtx=\LRFrxrad=1$ and $\Nfreq=2$.}
    \label{fig:analog_pattern}
\end{figure}

\section{Numerical Results}
\label{sec:numericalResults}

In this section, we numerically evaluate the proposed JCAS beamforming methods for both the hybrid MU-MIMO and the analog array architectures. In these numerical evaluations, we consider a realistic linear patch array simulated with CST Studio Suite, including the coupling effects between TX and RX elements to accurately model and investigate the SI phenomenon and its suppression. In particular, a linear array of $64$ elements is used, with ${\Ltx=32}$ TX and ${\Lrxrad=32}$ RX elements. The system's center frequency is assumed to be $28~\GHz$ while a nominal channel bandwidth of $500~\MHz$ is considered. %It is important to highlight that the same design principles can be applied to alternative array configurations \cite{heino_EuCAP2021}. 
Finally, while the true SI channel characteristics $\Hbold_{\SI, n}$ are provided by the electromagnetic simulations, we model the non-ideal estimation process through additive SI channel estimation error matrix $\tilde{\Hbold}_{\SI, n} \in \complexset{\Lrxrad}{\Ltx}$, expressed formally as
\begin{equation}
    \hat{\Hbold}_{\SI, n} = \Hbold_{\SI, n}+ \tilde{\Hbold}_{\SI, n}.
\end{equation}
In the numerical evaluations, we assume all the elements of $\tilde{\Hbold}_{\SI, n}$ to be independent complex Gaussian random variables, and the relative estimation error level is controlled by the parameter
\begin{equation}
\label{eq:SI_epsilon}
    \epsilon^2 =
    \frac{ \expectation \{  | \{ \tilde{\Hbold}_{\SI, n} \} _{l_\rx,l_\tx}
     |^2   \}}
    {\frac{1}{\Ltx \Lrxrad} 
    \sum_{l_\rx=1}^{\Lrxrad} \sum_{l_\tx=1}^{\Ltx}  
    | \{ \Hbold_{\SI, n} \} _{l_\rx,l_\tx}
     |^2 },
\end{equation}
for all $l_\rx$ and $l_\tx$.

\subsection{Hybrid MU-MIMO JCAS Architecture}

% Figures to show in the subsection
% \begin{itemize}
%     \item Fig.~\ref{fig:MIMO_patterns}\subref{fig:MIMO_TX} Effective TX pattern for each stream
%     \item Fig.~\ref{fig:MIMO_patterns}\subref{fig:MIMO_RX} Subarray's RX patterns
%     \item Fig.~\ref{fig:MIMO_muAnalysis} Radar gain vs. mu (Trade-off between communication and sensing)
% \end{itemize}

%Figure~\ref{fig:MIMO_patterns} shows illustrative examples of the TX and RX gain patterns for the
We first consider a hybrid MU-MIMO JCAS architecture when $\LRFtx=8$ and $\LRFrxrad=4$ RF chains are adopted at the TX and RX sides, respectively. Furthermore, we assume that $U=2$ and that the JCAS system provides two communication beams for the two users located at $\theta_{\comm,1}=-30^\circ$ and $\theta_{\comm,2}=30^\circ$, respectively, with the minimum TX gains of $\mu_1 = \mu_2 = 15~\dBi$ considered in the beamforming optimization in (\ref{eq:MIMO_constraint2}). Additionally, the TX power at the assumed radar direction of ${\theta_\rad=10^\circ}$ is maximized by generating a separate beam for sensing the environment, as described in Section~\ref{sec:MIMO_beamforming_optimization}.

The respective optimized beam-patterns are illustrated in Fig.~\ref{fig:MIMO_patterns}. Specifically, Fig.~\ref{fig:MIMO_patterns}\subref{fig:MIMO_TX} shows the effective TX patterns for each of the two transmitted streams. As it can be observed, each effective pattern provides two beams, one for communication and another one for sensing, while the IUI is minimized by imposing a null at the other user's direction as explained in (\ref{eq:MIMO_opt_problem}). For example, the $u=1$ stream provides two beams at $\theta_{\comm,1}=-30^\circ$ and ${\theta_\rad=10^\circ}$ while cancelling the $\theta_{\comm,2}=30^\circ$ direction.
Fig.~\ref{fig:MIMO_patterns}\subref{fig:MIMO_RX} shows the corresponding RX beam-patterns for the different RX RF chains (different subarrays). As it can be seen, each subarray connected to a particular RX RF chain generates a directive beam at the radar direction for observing the reflections. In addition, the RX weights are optimized to cancel the SI as will be explicitly illustrated in Section~\ref{sec:results_SIC}, namely Fig.~\ref{fig:SIC_MIMOcancellation} therein.

Figure~\ref{fig:MIMO_muAnalysis} illustrates and analyzes the effects of the minimum required TX gain imposed by the communication system into the achievable radar performance. In this case, we consider a similar scenario to Fig.~\ref{fig:MIMO_patterns} with $U=2$ communication users with $\mu_1 =\mu_2$. Moreover, different hybrid architectures are investigated by varying the number of TX RF chains. Note that the specific case of $\LRFtx=\Ltx=32$ corresponds already to full digital TX beamforming. The figure shows the behavior of the optimized radar gain, expressed in (\ref{eq:Power_radar_MIMO}), for different values of $\mu_u$. As can be observed, varying the parameter $\mu_u$ provides a trade-off between the communication and sensing performances. 
For relatively low values of $\mu_u$, the JCAS system allows more flexibility for sensing, providing higher gains at the radar direction. In contrast, when higher gains are required for communications beams (high values of $\mu_u$), the JCAS system will decrease the radar performance.
In addition, we can see how the different parametrizations of the hybrid architecture impact the final sensing performance. In this regard, the configurations with more TX RF chains and subsequently more flexibility in the digital domain TX processing, show better beamforming performance.

\subsection{Analog Array JCAS Architecture}

% Figures to show in the subsection
% \begin{itemize}
%     \item Fig.~\ref{fig:analog_pattern}\subref{fig:analog_TXpattern} TX pattern for analog 
%     \item Fig.~\ref{fig:analog_pattern}\subref{fig:analog_RXpattern} RX patterns for CF-A, CF-B, CF-C and PSL
%     \item Fig.~\ref{fig:analog_pattern}\subref{fig:analog_CRPpattern} CRP patterns for CF-A, CF-B, CF-C and PSL
% \end{itemize}

Next, we study and illustrate the beam-patterns in the analog array based JCAS architecture. Specifically, we pursue the multibeam operation with $U=2$ communication users located at $\theta_{\comm,1}=-40^\circ$ and $\theta_{\comm,2}=40^\circ$, respectively, together with one simultaneous sensing beam at ${\theta_\rad=10^\circ}$. The array sizes are as noted in the beginning of this section, namely ${\Ltx=\Lrxrad=32}$ while $\Nfreq=2$. 

The obtained beam-patterns are illustrated in Fig.~\ref{fig:analog_pattern}.
%Firstly, we analyze the beamforming performance of the proposed CF approaches when $\Nfreq=2$ frequency nulls is used to suppress the SI signal.
%
Specifically, at the TX side, the weights of the three considered beams are first separately optimized based on (\ref{eq:CF_TX_separate}) while then combined according to (\ref{eq:CF_TX_combination}) wherein we further assume $\rho_\rad=\rho_{\comm,1}=\rho_{\comm,2}=1/3$. The obtained corresponding TX beam-pattern is illustrated in Fig.~\ref{fig:analog_pattern}\subref{fig:analog_TXpattern} showing well-behaving beams at the three indicated directions. At the RX side, in turn, the two closed-form solutions called CF-B and CF-C are first explored, and illustrated in Fig.~\ref{fig:analog_pattern}\subref{fig:analog_RXpattern}. In the CF-B configuration, only SI suppression is pursued, in addition to providing coherent combining at the radar direction. In the more advanced CF-C configuration, in turn, additional angular nulls are considered in the RX side, as described in (\ref{eq:analog_CF_freqAngNulls}), in order to also suppress the possible negative effects of the communication beams on the overall sensing performance. In this case, two angular nulls ($\Nang=2$) are deployed at the corresponding communication directions.
As can be observed in Fig.~\ref{fig:analog_pattern}\subref{fig:analog_CRPpattern}, the CF-C configuration is able to suppress the communication beams contribution in the CRP.

%Secondly, we analyze a concrete example of 
While the CF-C configuration is able to impose explicit RX nulls at selected directions, the proposed CPSL-based beamforming optimization approach goes beyond this and seeks to suppress the overall clutter from different directions outside the sensing beam mainlobe. %, incorporating the same three beams. In this case, we assume 
We next illustrate the power of such approach in the same ABF JCAS scenario, while assume a desired response of $\tilde{G}^\RF_{\CRP,\maxText}=35~\dBi$, $\Delta=14^\circ$ and $CPSL=-75~\dB$ as shown in Fig.~\ref{fig:analog_pattern}\subref{fig:analog_CRPpattern}.
As it can be observed, the proposed CPSL optimization which provides more flexibility in the beamforming design, shows the best performance in terms of mitigating or suppressing all possible interference sources outside the sensing mainlobe with a CPSL of around $-75~\dB$. In contrast with the closed-form solution, the main advantage of this approach is the capability of controlling the CPSL for the non-radar directions, and therefore reduce the interference due to the communication beams.

\vspace{-3mm}
\subsection{Self-Interference Suppression Analysis}
\label{sec:results_SIC}

% Figures to show in the subsection
% \begin{itemize}
%     \item Fig.~\ref{fig:SIC_MIMOcancellation} Example of SI frequency channel for Hybrid MU-MIMO JCAS architecture
%     \item Fig.~\ref{fig:SIC_numberNulls} Frequency spectrum for different number of nulls using analog array JCAS architecture
%     \item Fig.~\ref{fig:SIC_epsilon}\subref{fig:SIC_epsilon_cancellation} Cancellation vs. SI estimation error
%     \item Fig.~\ref{fig:SIC_epsilon}\subref{fig:SIC_epsilon_RXgain} RX gain vs. SI estimation error
% \end{itemize}

\begin{figure}[t]
        \centering
        \includegraphics[width=1\columnwidth]{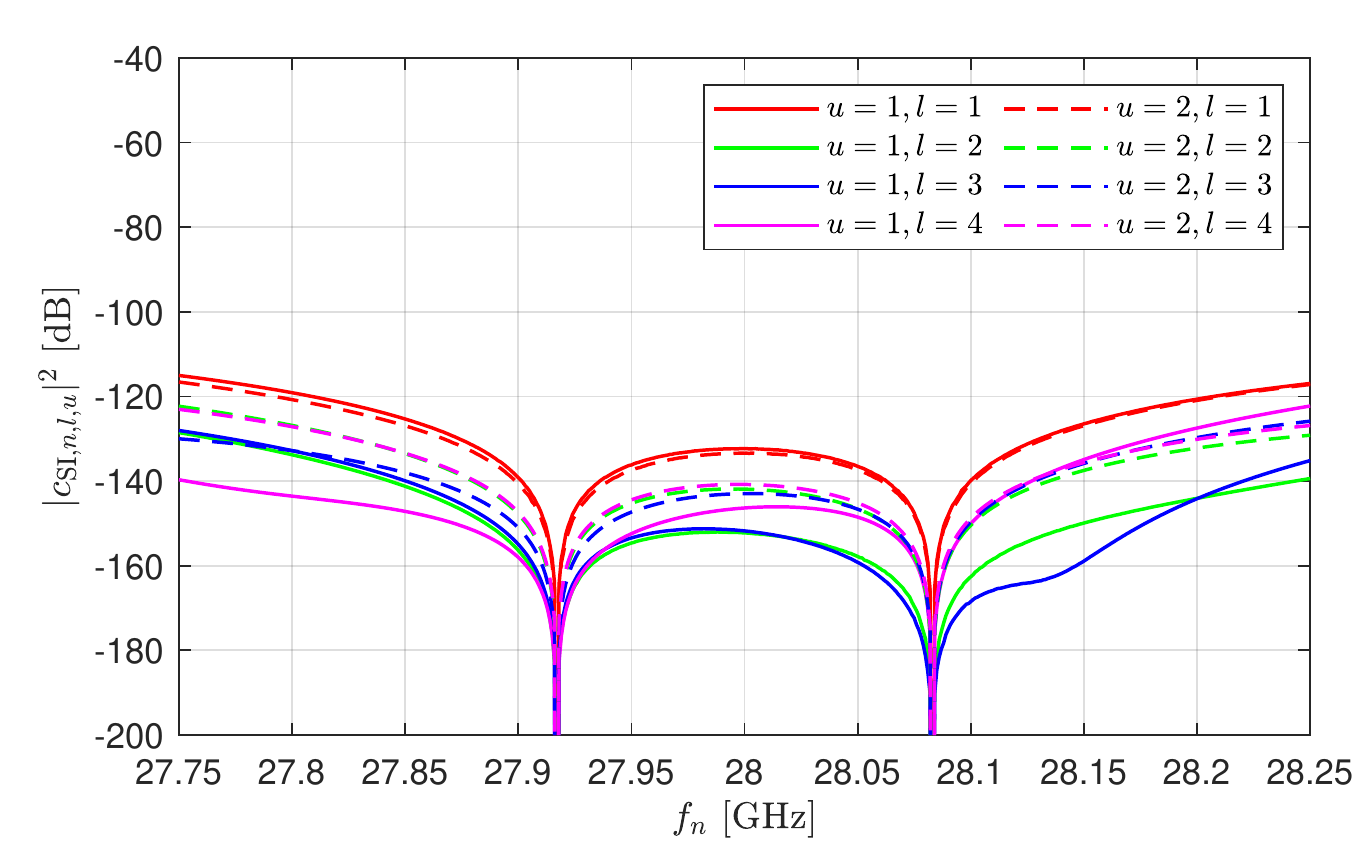}
        \caption{Illustration of the SI suppression performance in hybrid MU-MIMO JCAS case, showing the effective SI channel between the $\thh{u}$ TX stream and the $\thh{l}$ RX RF chain. The considered system parameters are: ${\theta_\rad=10^\circ}$, $U=2$, $\theta_{\comm,1}=-30^\circ$, $\theta_{\comm,2}=30^\circ$, ${\Ltx=\Lrxrad=32}$, $\LRFtx=8$, $\LRFrxrad=4$, $\mu_1 = \mu_2 = 15~\dBi$, and $\Nfreq=2$.  
        }
        \label{fig:SIC_MIMOcancellation}
\end{figure}

\begin{figure}[t]
        \centering
        \includegraphics[width=1\columnwidth]{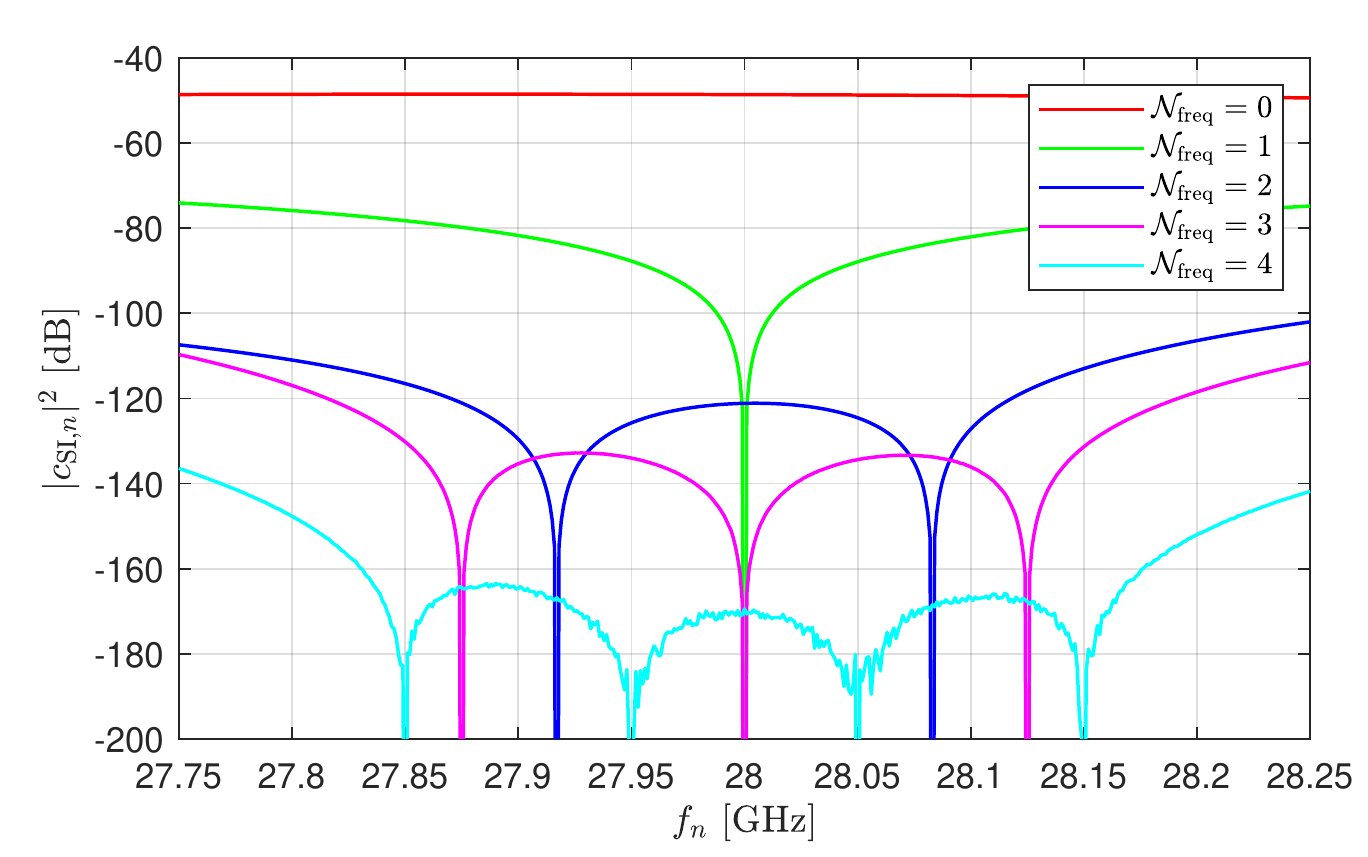}
        \caption{Illustration of the SI suppression performance for different numbers of the frequency nulls ($\Nfreq$) in the analog array JCAS case using the CF-B configuration. The considered system parameters are: $U=2$, ${\theta_\rad=10^\circ}$, $\theta_{\comm,1}=-40^\circ$, $\theta_{\comm,2}=40^\circ$ and ${\Ltx=\Lrxrad=32}$. 
        The CF-A configuration is used as a reference and corresponds to $\Nfreq=0$.
        }
        \label{fig:SIC_numberNulls}
\end{figure}

\begin{figure}[!t]
    \centering
        \subfloat[Average SI suppression performance]{
            \begin{tikzpicture}
                \draw (0, 0) node[inner sep=0]{
                    \includegraphics[width=1\columnwidth]{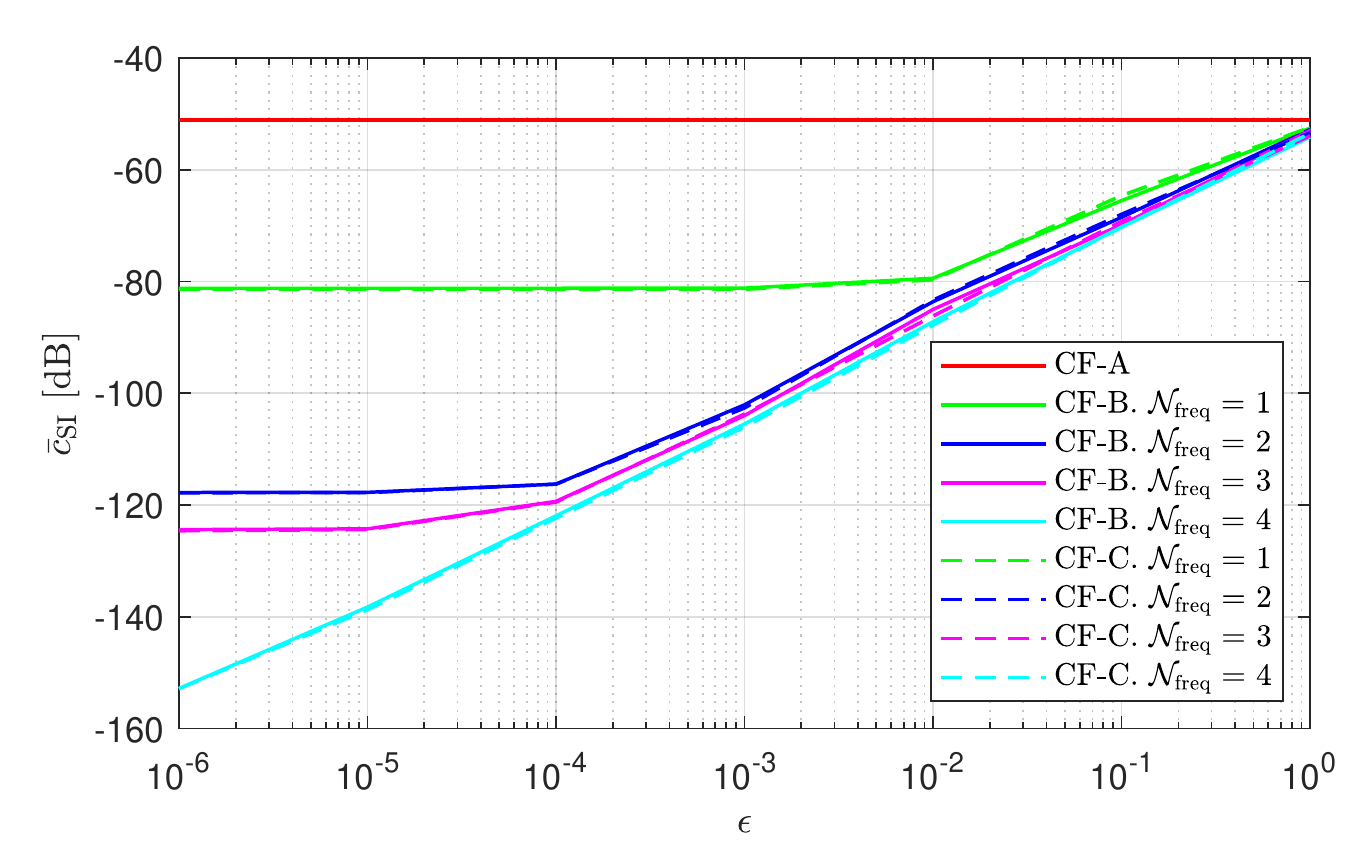}
                    \label{fig:SIC_epsilon_cancellation}
                };
                \draw (3.1, 0.41) node {{\scriptsize \cite{journal_7}}};
            \end{tikzpicture}
        }
        \hfil
        \subfloat[CRP gain]{
            \begin{tikzpicture}
                \draw (0, 0) node[inner sep=0]{
                    \includegraphics[width=1\columnwidth]{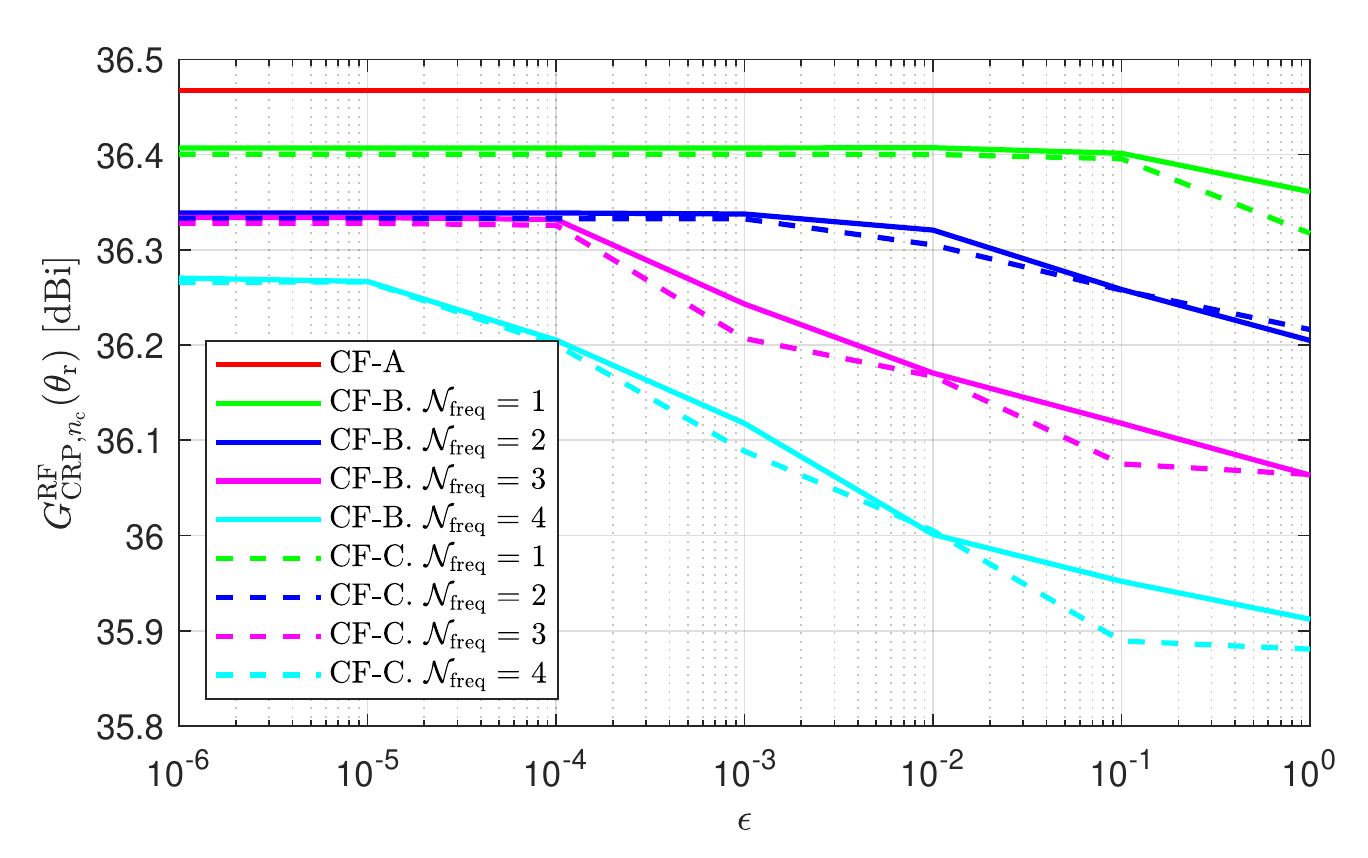}
                    \label{fig:SIC_epsilon_RXgain}
                };
                \draw (-1.6, 0.41) node {{\scriptsize \cite{journal_7}}};
            \end{tikzpicture}
        }
    \caption{Illustration of (a) the SI suppression performance and (b) the beamforming performance as functions of the relative SI channel estimation error level, $\epsilon$, in the analog array JCAS case. Also different numbers of the frequency nulls ($\Nfreq$) are considered and shown. The main system parameters are: $U=2$, ${\theta_\rad=10^\circ}$, $\theta_{\comm,1}=-30^\circ$, $\theta_{\comm,2}=30^\circ$ and ${\Ltx=\Lrxrad=32}$.}
    \label{fig:SIC_epsilon}
    \vspace{-4mm}
\end{figure}

% \begin{figure}[!t]
%     \centering
%         \subfloat[Average SI suppression performance]{\includegraphics[width=1\columnwidth]{figures/SIC_epsilon_cancellation_v1.pdf}
%         \label{fig:SIC_epsilon_cancellation}}
%         \hfil
%         \subfloat[CRP gain]{\includegraphics[width=1\columnwidth]{figures/SIC_epsilon_RXgain_v1.pdf}
%         \label{fig:SIC_epsilon_RXgain}}
%     \caption{Illustration of (a) the SI suppression performance and (b) the beamforming performance as functions of the relative SI channel estimation error level, $\epsilon$, in the analog array JCAS case. Also different numbers of the frequency nulls ($\Nfreq$) are considered and shown. The main system parameters are: $U=2$, ${\theta_\rad=10^\circ}$, $\theta_{\comm,1}=-30^\circ$, $\theta_{\comm,2}=30^\circ$ and ${\Ltx=\Lrxrad=32}$.}
%     \label{fig:SIC_epsilon}
% \end{figure}

We next address and analyze the SI suppression performance that can be achieved through the NSP beamforming solutions.
Figure \ref{fig:SIC_MIMOcancellation} illustrates the obtained SI suppression behavior as a function of frequency in the hybrid MU-MIMO JCAS architecture case when design parameters similar to Fig.~\ref{fig:MIMO_patterns} are utilized. Here, ideal SI channel estimation capability is yet considered, implying that $\epsilon = 0$. Specifically, the figure illustrates the effective frequency-domain characteristics of the beamformed SI channel in (\ref{eq:NSP_effectiveSIchannel_vector}) between the $\thh{u}$ TX stream and the $\thh{l}$ RX RF chain when $\Nfreq=2$ frequency nulls are deployed in the beamforming optimization. As it can be observed, each RX RF chain is subject to different effective SI channels. More concretely, the RX RF chain with $l=1$ whose subarray is the closest to the TX array exhibits most severe SI coupling. However, overall, the obtained SI suppression numbers are excellent over the whole channel bandwidth.

Figure \ref{fig:SIC_numberNulls} shows the frequency-domain beamformed SI channel characteristics in the corresponding analog array JCAS architecture case, when design parameters similar to Fig.~\ref{fig:analog_pattern} are used. In this case, we focus on the CF-B beamformer design approach and analyze the effects of the number of frequency nulls, $\Nfreq$, on the SI suppression. Additionally, the special case of $\Nfreq=0$ corresponds to CF-A which serves as a reference or baseline.
In CF-B design, the frequency nulls are chosen such that they are uniformly distributed within the whole bandwidth, as $f_{n'} = 27.75 + 0.5 \frac{n'}{\Nfreq+1}~\GHz$ with $n'=1,2,\ldots,\Nfreq$.
As can be observed through the results, increasing the number of frequency nulls clearly improves the SI suppression. 
The case of $\Nfreq=1$ provides an average SI suppression within the considered frequency channel of around $-80~\dB$, while if the number of nulls is increased to $\Nfreq=4$ we can achieve a SI suppression level of around $-160~\dB$ already. We note that the baseline level of around $-50~\dB$ even when no beamforming based SI suppression is applied ($\Nfreq=0$) stems from the the basic physical isolation between the TX and RX antenna arrays.

% \begin{figure*}[t]
%     \centering
%         \subfloat[Sensing scenario]{\includegraphics[width=0.165\textwidth]{figures/sensing_scenario.pdf}
%         \label{fig:sensing_scenario}}
%         \hfil
%         \subfloat[CF-A]{\includegraphics[width=0.19\textwidth]{figures/sensing_CF_A_v1.pdf}
%         \label{fig:sensing_CF_A}}
%         \hfil
%         \subfloat[CF-B]{\includegraphics[width=0.19\textwidth]{figures/sensing_CF_B_v1.pdf}
%         \label{fig:sensing_CF_B}}
%         \hfil
%         \subfloat[CF-C]{\includegraphics[width=0.19\textwidth]{figures/sensing_CF_C_v1.pdf}
%         \label{fig:sensing_CF_C}}
%         \hfil
%         \subfloat[CPSL optimization]{\includegraphics[width=0.19\textwidth]{figures/sensing_PSL_v1.pdf}
%         \label{fig:sensing_PSL}}
%     \caption{(a) Considered scattering scenario with $20$ targets and $U=2$ communication users located at $\theta_{\comm,1}=-40^\circ$ and $\theta_{\comm,2}=40^\circ$ for different analog array JCAS configurations (b) CF-A, (c) CF-B, (d) CF-C and (e) CPSL optimization, respectively. The analog array JCAS system senses the environment from $-60^\circ$ to $-60^\circ$ with a step of $1^\circ$.
%     }
%     \label{fig:sensing_comparison}
% \end{figure*}

\begin{figure*}[t]
    \centering
        \subfloat[Sensing scenario]{\includegraphics[width=0.3\textwidth]{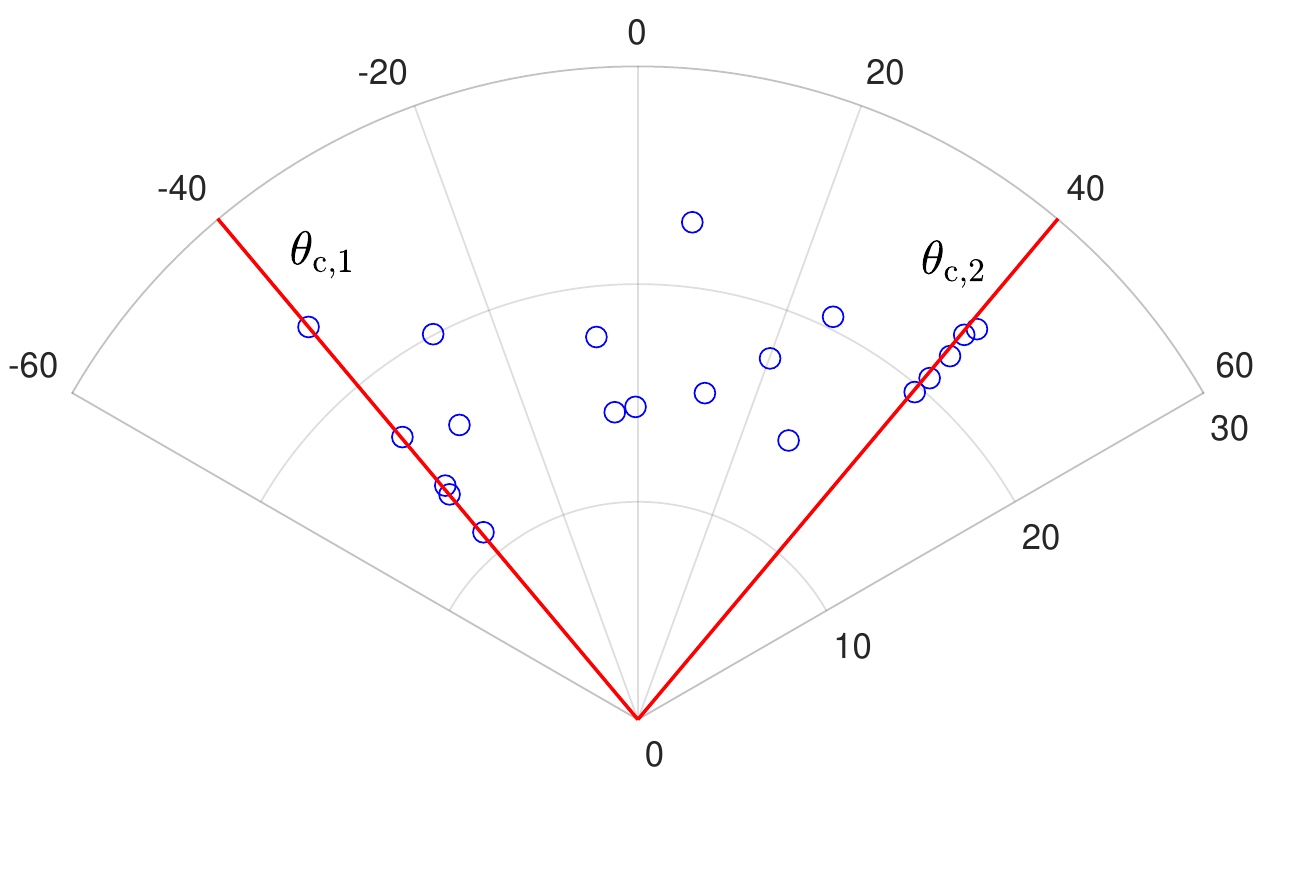}
        \label{fig:sensing_scenario}}
        \hfil
        \subfloat[Analog array, CF-A \cite{journal_7}]{\includegraphics[width=0.3\textwidth]{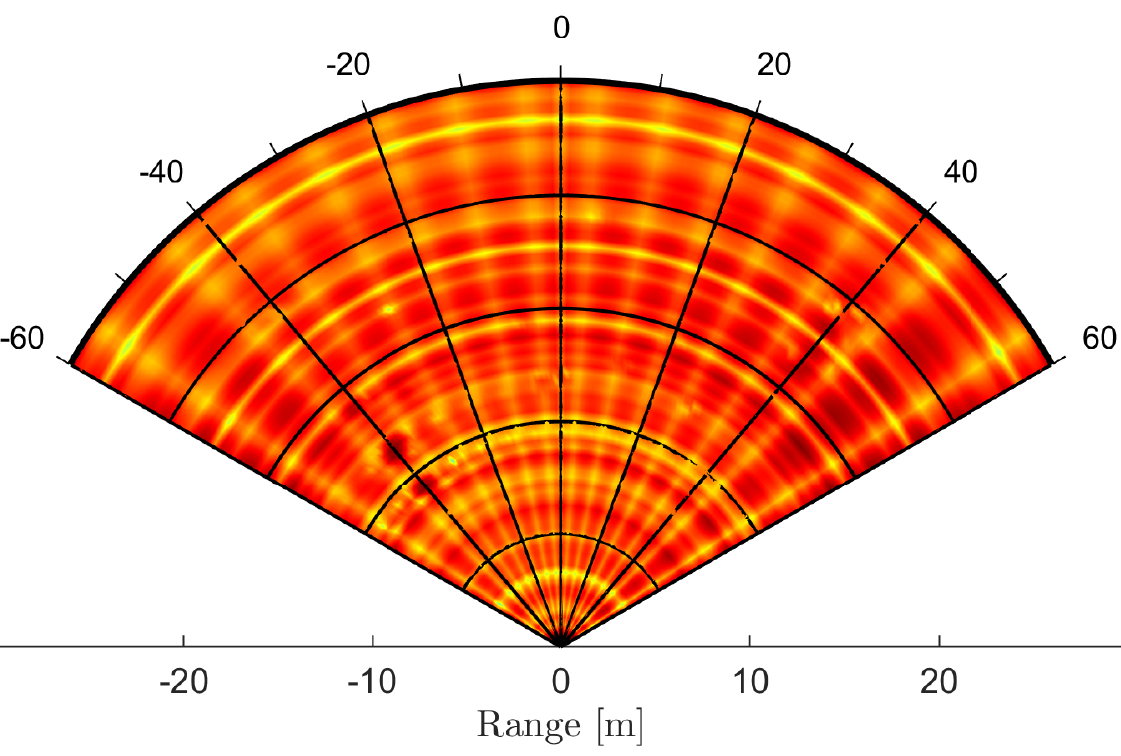}
        \label{fig:sensing_CF_A}}
        \hfil
        \subfloat[Analog array, CF-B]{\includegraphics[width=0.3\textwidth]{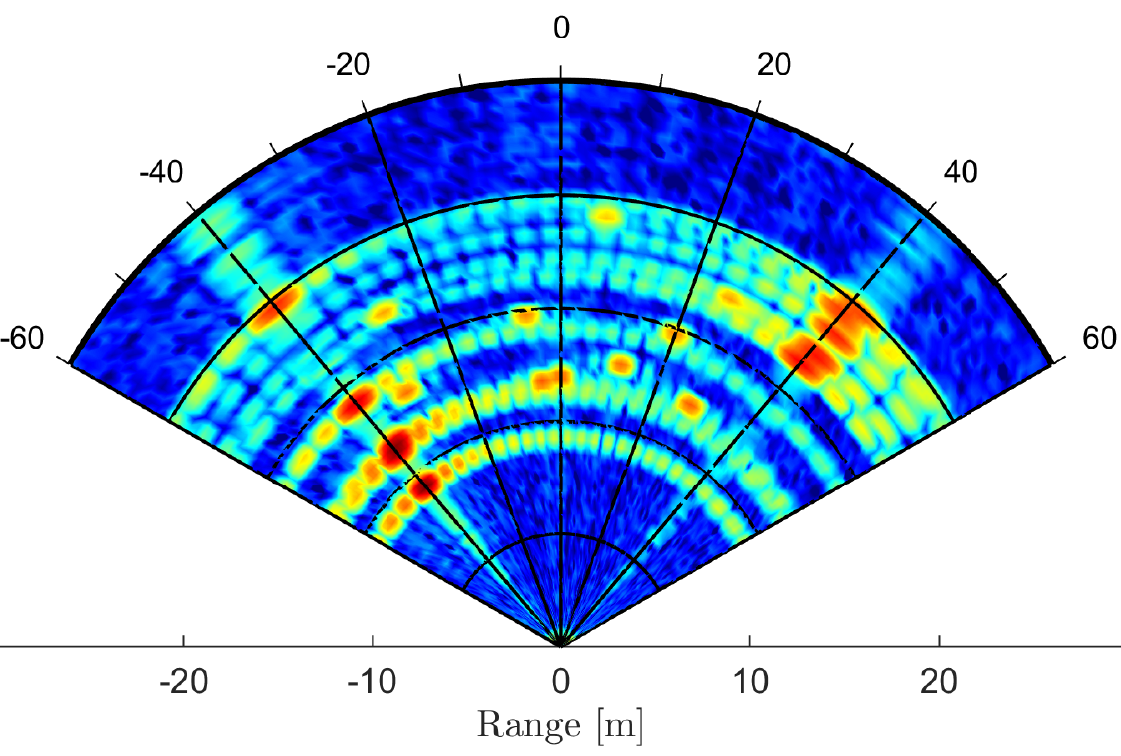}
        \label{fig:sensing_CF_B}}
        \hfil
        \\
        \subfloat[Analog array, CF-C]{\includegraphics[width=0.3\textwidth]{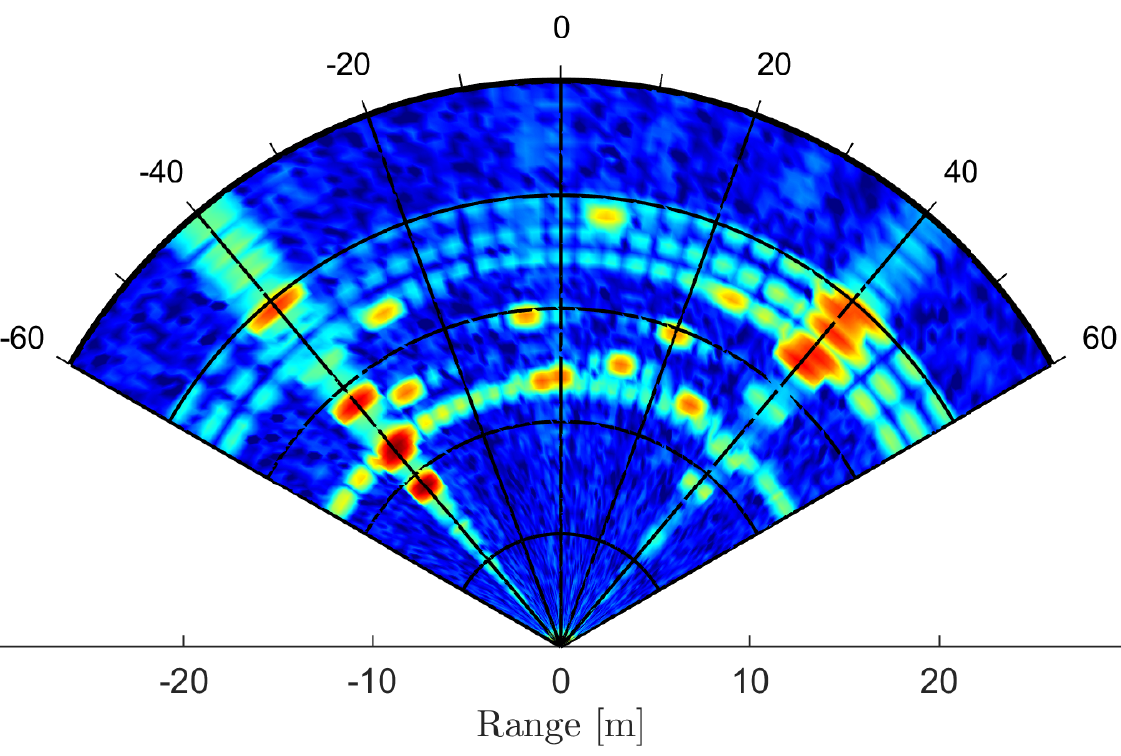}
        \label{fig:sensing_CF_C}}
        \hfil
        \subfloat[Analog array, CPSL optimization]{\includegraphics[width=0.3\textwidth]{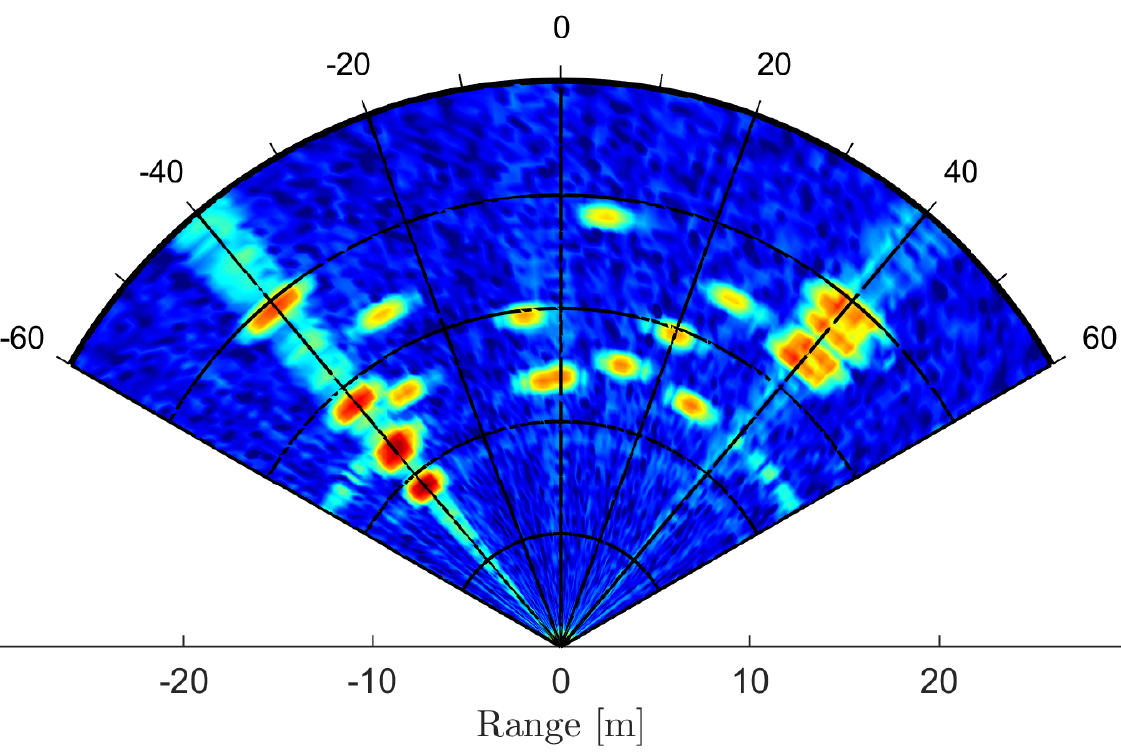}
        \label{fig:sensing_PSL}}
        \hfil
        \subfloat[Hybrid MU-MIMO]{\includegraphics[width=0.3\textwidth]{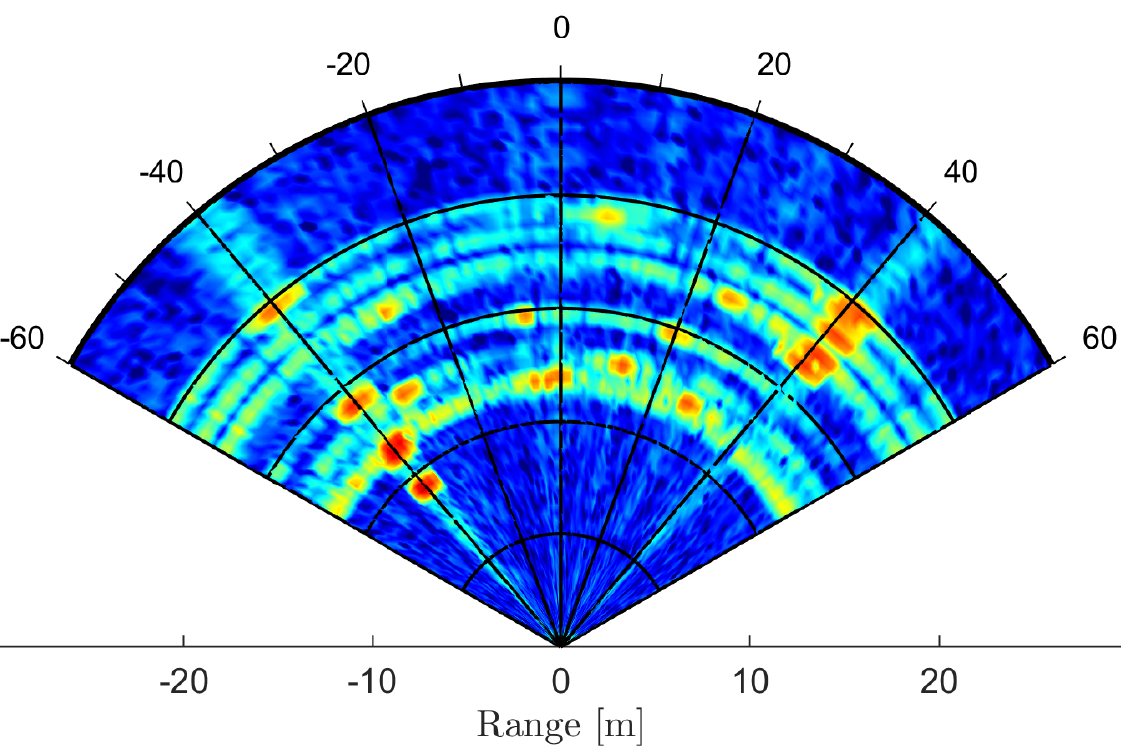}
        \label{fig:sensing_MIMO}}
    \caption{In (a), the considered scattering environment with $20$ targets and $U=2$ communication users located at $\theta_{\comm,1}=-40^\circ$ and $\theta_{\comm,2}=40^\circ$ is illustrated. The corresponding sensing results in the analog array JCAS case with different beamforming configurations are shown, as follows: (b) CF-A, (c) CF-B, (d) CF-C and (e) CPSL optimization. Moreover, (f) shows the sensing results using the hybrid MU-MIMO JCAS architecture for the same scenario. For all the considered architectures, the JCAS system senses the environment from $-60^\circ$ to $-60^\circ$ with a step of $1^\circ$.
    }
    \label{fig:sensing_comparison}
\end{figure*}

We next assess how imperfect SI channel estimates impact the achievable SI suppression performance in the context of the presented beamforming schemes. %are subject to imperfect SI channel estimation (\ref{eq:SI_epsilon}). 
To this end, Fig.~\ref{fig:SIC_epsilon} illustrates how the level of the SI channel estimation error, parametrized by $\epsilon$ in (\ref{eq:SI_epsilon}), degrades different system metrics when deploying the  alternative beamforming configurations summarized in Table~\ref{tab:CF_configurations}. As shown in Fig.~\ref{fig:SIC_epsilon}\subref{fig:SIC_epsilon_cancellation}, the achievable SI suppression is closely tied with the estimation accuracy. By design, the considered configurations are able to achieve different SI suppression levels depending on the number of nulls when accurate SI channel estimates available. 
On the other hand, as expected, when the SI channel estimation accuracy decreases, the SI suppression essentially converges to the CF-A configuration ($\Nfreq=0$, only physical isolation), independent of the value of $\Nfreq$. It is also noted that the \mbox{CF-C} configuration, which includes additional $\Nang$ angular nulls to suppress the impact of the communication beams, shows a similar performance compared to the CF-B configuration in terms of SI suppression and its sensitivity to SI channel estimation accuracy.

Furthermore, since the SI channel estimation error directly affects the calculation of the NSP matrix in (\ref{eq:NSP_multipleFrequencies}), it also impacts the RX beamforming performance. Fig.~\ref{fig:SIC_epsilon}\subref{fig:SIC_epsilon_RXgain} illustrates the degradation of the CRP radar gain for different SI channel estimation error levels. 
Specifically, it can be observed that the number of frequency nulls used for the SI suppression just produces an insignificant CRP loss of around $0.5~\dB$, which does not compromise the radar performance of the proposed JCAS system.
%
%directly affects the achieved CRP gain. %, regardless of the SI channel estimation error. 
%However, this CRP degradation of less than $1~\dB$ is insignificant compared to the losses in the two-way radar propagation and does not compromise the radar performance of the proposed JCAS system.
In addition, if we compare mutually the CF-B and CF-C design approaches, we observe how the angular nulls included in CF-C slightly degrade the CRP gain performance. This performance degradation, due to the loss of beamforming flexibility when applying the presented NSP method, results in a trade off between the desired SI suppression, CRP gain and the ability to suppress communication beams at the RX side.

% Specifically, it can be observed that the number of frequency nulls used for the SI suppression directly affects the achieved CRP gain, regardless of the SI channel estimation error. In addition, if we compare mutually the CF-B and CF-C design approaches, we observe how the angular nulls included in CF-C slightly degrade the CRP gain performance. This performance degradation, due to the loss of beamforming flexibility when applying the presented NSP method, results in a trade off between the desired SI suppression, CRP gain and the ability to suppress communication beams at the RX side.

\subsection{Representative Sensing Results}

%\blue{Include also sensing results for hybrid MU-MIMO JCAS case?}
% Figures to show in the subsection
% \begin{itemize}
%     \item Fig.~\ref{fig:sensing_comparison} Sensing for CF-A, CF-B, CF-C and PSL
%     \item Some results as well for MIMO? DOA and range estimation?
% \end{itemize}

Finally, we present example numerical results to validate and assess the actual sensing capability with the different JCAS architectures and proposed beamforming methods, in a scattering environment with 20 static point targets. 
Ten of these targets are deliberately placed in the direction of the communication beams, at $\theta_{\comm,1}=-40^\circ$ and $\theta_{\comm,2}=40^\circ$, with a radar cross section (RCS) of $10~\meter^2$ as shown in Fig.~\ref{fig:sensing_comparison}\subref{fig:sensing_scenario}. 
The rest of the targets are uniformly distributed in the sensed area at distances within $10~\meter$ to $25~\meter$ and angles from $-40^{\circ}$ to $40^{\circ}$, with RCS of $1~\meter^2$. The transmission power is $20~\dBm$ and the total thermal noise in the receiver is $-88~\dBm$. 
The sensing beam scans the angles between $-60^{\circ}$ to $60^{\circ}$ with a step of $1^{\circ}$. 
According to \cite{3GPPTS38104}, we consider an OFDM waveform with $N=3168$ subcarriers, $10$ symbols and $\SCS=120~\kHz$ providing a transmission bandwidth of around $400~\MHz$.
For the actual radar processing to construct the range--angle radar images, we adopt subcarrier-domain processing that utilizes directly the TX and RX subcarrier samples similar to \cite{thesis_braun2014, conference_12, conference_13,myPaper_2_TMTT19}, including also a Hamming window to control the sidelobe level. Furthermore, to be able to focus on the impacts of the co-existing communications and sensing beams, we assume for simplicity that ideal SI channel estimates are available. % estimation process of the SI response with $\epsilon = 0$.

First, in Fig.~\ref{fig:sensing_comparison}\subref{fig:sensing_CF_A}, we illustrate the CF-A beamforming case meaning that no SI nor communication beam suppression is applied. It is noted that in these numerical simulations, we assume sufficiently large RX dynamic range to observe both the target reflections and the strong SI component. However, in practice, this beamforming configuration will easily lead to the saturation of the RX chain, preventing any subsequent sensing functionality. In Fig.~\ref{fig:sensing_comparison}\subref{fig:sensing_CF_A}, we can observe how the strong SI appears as a strong close-by target whose sidelobes are spread in the range profiles, masking possible weak targets.
Then, the same scenario is assessed with the CF-B beamforming configuration as shown in Fig.~\ref{fig:sensing_comparison}\subref{fig:sensing_CF_B}. In this case, the strong SI is suppressed by the NSP method including $\Nfreq=2$ frequency nulls. However, it can be clearly observed that the communication beam interference still generates strong sidelobes, which produce a substantial masking effect along the rest of the sensing directions, thus potentially preventing the detection of weak targets. To overcome this problem, the communication beam is suppressed in the RX side by incorporating angular nulls at the communication directions using the beamforming configuration CF-C, with the corresponding results shown in Fig.~\ref{fig:sensing_comparison}\subref{fig:sensing_CF_C}. In this case, the reflections due to the communication beam are suppressed to a certain extent, already improving the sensing capability.

Furthermore, Fig.~\ref{fig:sensing_comparison}\subref{fig:sensing_PSL} shows the radar image when the proposed CPSL optimization method is implemented. By applying the proposed CPSL optimization based beamforming design, the sidelobe levels can be largely suppressed allowing thus to efficiently avoid such masking problem. 
As can be observed, the radar image is still clearly improved compared to the different variants of the closed-form method in Figs.~\ref{fig:sensing_comparison}\subref{fig:sensing_CF_A}--\subref{fig:sensing_CF_C}. Specifically, all the targets are efficiently visible as can be concluded by comparing the subfigures (a) and (e).

Finally, we sense the same scenario with the hybrid MU-MIMO JCAS architecture by using the beamformer design presented in Section~\ref{sec:MIMO_beamforming}. The corresponding radar image is visually shown in Fig.~\ref{fig:sensing_comparison}\subref{fig:sensing_MIMO}. In this case, we utilize a basic radar processing approach, where the received signals at different RX RF chains are combined to maximize the received power at the radar direction while the communication directions are suppressed similar to the CF-C configuration. However, this hybrid architecture, which provides multiple parallel digital signals, will also support more advanced radar processing and related techniques, e.g., multiple signal classification (MUSIC), that will improve the direction of arrival estimation performance \cite{conference_40, myPaper_8_JCS21}.

\section{Conclusions}
\label{sec:Conclusions}

In this article, the joint communication and sensing paradigm was addressed and studied, with special focus on the fundamental challenges with the transmitter--receiver isolation and the co-existing communications and sensing beams. % beamforming design challenges as the beamforming and self-interference cancellation for JCAS operation. 
We considered both analog beamforming and hybrid analog--digital beamforming based JCAS systems, and proposed multiple new beamformer design and optimization solutions accordingly, addressing the noted challenges with the TX-RX isolation and the co-existing beams. 
Specifically, in the context of hybrid beamforming based MU-MIMO JCAS systems, we formulated and solved TX and RX beamforming optimization problems that maximize the beamformed power at the sensing direction while constraining the beamformed power at the communications directions, cancel the IUI and suppress the wideband SI by implementing multiple frequency nulls. 
Additionally, for analog beamforming based JCAS systems, alternative TX and RX beamforming solutions were proposed, which allow the integration of multiple beams for communications and sensing while suppressing the SI. In particular, we observed that the optimization of the combined radar pattern enhances the beamforming radar performance in the analog JCAS context.
Finally, we analyzed the performance of the proposed methods through extensive simulations with a realistic linear patch array, showing that substantial gains and benefits can be achieved with the proposed beamforming and SI cancellation techniques.

\bibliographystyle{IEEEtran}
% \bibliography{mainReferences}
\bibliography{main}

\vfill

\end{document}